\documentclass[aip,amsmath,amssymb,reprint,jcp]{revtex4-1}

\usepackage[version=3]{mhchem} 
\usepackage{subfigure}
\usepackage{graphicx}
\usepackage{placeins}
\usepackage{titlesec} 
\usepackage{multirow}

\usepackage{makecell}

\begin{document}

\title[ESTEEM]
  {Predicting Spectroscopic Properties of Solvated Nile Red with Automated Workflows for Machine Learned Interatomic Potentials}

\author{Jacob Eller}
\affiliation{Department of Physics, University of Warwick}
\author{Nicholas D. M. Hine}
\email{n.d.m.hine@warwick.ac.uk}
\affiliation{Department of Physics, University of Warwick}

\renewcommand{\theparagraph}{\roman{paragraph}}
\titleformat{\paragraph}[runin]
  {\itshape}  
  {\theparagraph)}                   
  {0.5em}                            
  {}                                 
  [:\,\,]                            

\renewcommand{\thesubsubsection}{}

\maketitle

\section*{Abstract}
Machine Learned Interatomic Potentials (MLIPs) offer a powerful combination of abilities for accelerating theoretical spectroscopy calculations utilising both ensemble sampling and trajectory post-processing for inclusion of vibronic effects, which can be very challenging for traditional ab initio MD approaches. We demonstrate a workflow that enables efficient generation of MLIPs for the solvatochromic dye nile red system, in a variety of solvents. We use iterative active learning techniques to make this process as efficient as possible in terms of number and size of Density Functional Theory (DFT) calculations. Additionally, we compare the efficacy of various methodologies: generating distinct MLIPs for each adiabatic state, using one ground state MLIP in combination with delta-ML of excitation energies, and using a three-headed multiheaded ML model. To evaluate the validity of the resulting models, we compare predicted absorption and emission spectra to experimental spectra. We found that the incorporation of larger solvent systems into training data, and the use of delta models to predict the excitation energies, enables the accurate and affordable prediction of UV-Vis spectra with accuracy equivalent to the ground truth method, which is time-dependent DFT in this case. 

\section{Introduction}
The high degree of complexity and computational effort associated with multiscale modelling means that robust and transferable computational workflows are key to success.
A prime example is ab initio prediction of spectroscopic properties of solvated molecules.
Simulations which are fast and affordable, but incorporate the effects of the environment\cite{Nemkovich2002,Tomin1981} with high accuracy, enable routine prediction of structure-driven dynamical processes\cite{STIEL2001239,Ogunyemi2025} and their influence on spectroscopic features.
This can then significantly enhance how closely structure-function relations can be investigated.
In this work we present an overview and selected case studies for a package addressing the need for robust workflows in theoretical spectroscopy: the Explicit Solvent Toolkit for Electronic Excitations of Molecules (ESTEEM).
ESTEEM has been used previously in the context of full ab initio electronic structure calculations\cite{10.1063/1.4979196,doi:10.1021/acs.jpca.8b11013}.
Here, we present its features for training, testing and and utilising Machine Learned Interatomic Potentials (MLIPs) to reproduce the results of electronic structure theory calculations.

The motivation for a workflow management package for excitations in solvent is the need to balance competing demands of long timescales, high accuracy, and efficient calculation within reasonable walltime.
Ensemble sampling over long timescales is required to include both the reorganisation of the solvent shell and the dynamics of the solute and solvent molecules\cite{10.1063/1.5006043,Zuehlsdorff2016}.
The use of MLIPs enables efficient sampling of the medium-timescale dynamics of the ensemble via molecular dynamics\cite{https://doi.org/10.1002/advs.202409009,CASTELLANO2025109730}.
Longer-timescale sampling often requires fast empirical potentials\cite{https://doi.org/10.1002/cphc.201000109,RevModPhys.34.239}, but these are not sufficiently accurate\cite{https://doi.org/10.1002/qua.25719,chemistry5010004} for the short-timescale dynamics required not just by advanced spectroscopies such as pump-probe techniques\cite{Khitrova:88,CHERGUI2012398}, but also in accurate prediction of absorption and emission spectra\cite{Vogt2023}.
In this work we focus on absorption and emission spectroscopy, since these provide a straightforward yet stringent test of accurate prediction of solvent effects.
The challenge arises because ML predictions for electronic excitation spectra require representations of ground and excited state potential energy surfaces (PES) capable of high accuracy well away from their respective relaxed geometries.
For example, a vertical de-excitation contributing to an emission spectrum may involve a transition at a geometry near a minimum on the excited state PES to a point on the ground state PES that may be far from its minimum energy geometry.
Iterative procedures are therefore required to ensure adequate sampling of training data for an MLIP in such regions\cite{MOSQUEIRAREY2021553}, as training data sampled from configurations near minima on the ground state PES does a poor job of sampling near minima of the excited state PES\cite{D5SC05579B}.
Secondly, this variation of geometry precludes many WF-based methods which require the user to define the active space, as this can become highly problematic and perform inconsistently across widely-varying regions of phase space\cite{10.1063/1.4922352,10.1063/1.481132}.
Overall, the requirement is for a fast and efficient way to sample widely-varying points on the ground and excited PES, from which MLIP training data is obtained.
The underlying electronic structure method used must be sufficiently accurate to provide well-converged energy differences between the states, but also quick enough to apply to thousands of snapshots of clusters containing both the solute and nearby solvent molecules.

In this work, Density Functional Theory (DFT) \cite{PhysRev.136.B864,PhysRev.140.A1133} and time-dependent DFT (TD-DFT) \cite{PhysRevLett.52.997,casida2009time} are used to carry out ground and excited state calculations respectively.
%
%
%
%
%
%
%
The use of TD-DFT with hybrid functionals to predict spectroscopic properties of solutions is widely supported in literature, with the method appearing to capture much of the effect of the solvent environment\cite{doi:10.1021/jp100642c,doi:10.1021/acs.jpca.1c06621,ADAMO2000152}.


The requirement of large system sizes arises from the need to include sufficient range of solvent shells to account for the full effect of the solvent\cite{Zuehlsdorff2016}. 
While implicit solvent techniques can successfully describe some aspects of the effect of solvent on vertical electronic transitions, such as redshifts arising from screening of the transition dipole moment by the solvent, other effects such as hydrogen bonding and pi-pi stacking are in general beyond their abilities\cite{10.1063/5.0038342}.
Machine learned interatomic potentials (MLIPs) trained on a relatively small amount of highly accurate data offer a means to perform calculations
that include all relevant solvent-solute interactions, with relatively low computational expense.


When propagating long length- and time-scale dynamics, the accurate treatment of long-range interactions--such as dispersion and coulombic force--is crucial to ensure the physical traversal of phase space.
Thus, approaches have been developed to offer improved treatment of long-range interactions compared to the original `short-range' MLIPs.
One such technique involves including long-range corrections explicitly, by using a fixed functional form to calculate interactions between point charges further apart than the cutoff distance\cite{doi:10.1021/acs.jctc.9b00181}.
The resulting energies are summed on top of `short-range' MLIP predictions.
Another approach entails models implicitly learning dispersion forces by utilizing methodologies with dispersion corrections to obtain training data\cite{doi:10.1073/pnas.1602375113}.
Message-passing graph neural networks (MP-GNN) are widely considered the current state-of-the-art architecture in the context of MLIPs\cite{pmlr-v70-gilmer17a}.
MP-GNNs facilitate the exchange of messages communicating the properties/features of individual nodes (representing atoms) to neighboring nodes within a specific distance through multiple iterations of message passing.
This ultimately updates the features of nodes to contain information about their surrounding environment, with successive iterations further expanding the receptive field.
If systems used for training are large enough to capture long-range effects, and the receptive field of the model encapsulates these interactions, models can implicitly learn such interactions.



In this work, we outline the core methodology of ESTEEM in section \ref{sec:overview}, then demonstrate its effectiveness in section \ref{sec:nilered} with an attempt to predict ultraviolet-visible (UV-Vis) absorption and emission spectra of the fluorescent dye Nile Red in various solvents. 

\section{ESTEEM Overview}
\label{sec:overview}

\begin{figure*}
    \centering
    \includegraphics[width=1\linewidth]{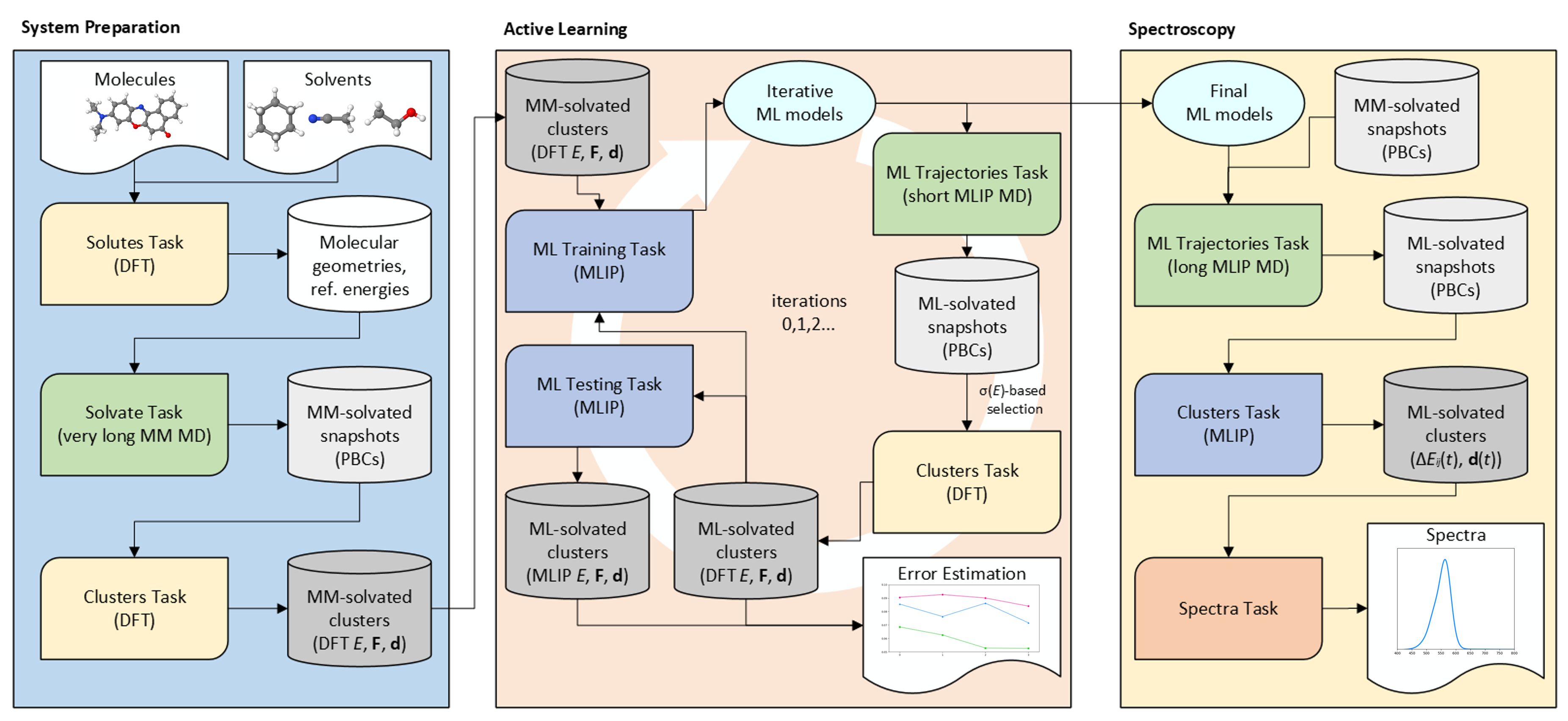}
    \caption{Visualisation of workflow in ESTEEM. Rounded rectangles represent ESTEEM's `tasks', cylinders represent datasets produced, ovals are trained ML models, and illustrative pages represent input choices and output data. The iterative loop within the Active Learning section is repeated 2-3 times, with selection of a set of new geometries for DFT calculations based on evaluating the standard deviation of energy predictions for a cluster over a committee of ML models.}
    \label{fig:esteem_workflow}
\end{figure*}

Use of ESTEEM proceeds via execution of a series of tasks defined by an input script, which specifies the parameters of each task, and defines a listing of solute and solvent molecules to iterate over. Each task typically utilizes an external software package to perform the main body of the task, preceded and followed by pre- and post-processing with Python. 
Different external packages will have different strengths and weakness for different target systems, so the ability to easily switch between such packages while maintaining a workflow is vital.
ESTEEM enables this by providing wrappers for a range of packages, allowing the user to switch them by simply assigning a different wrapper as a parameter of a task, or a new wrapper can straightforwardly be written, such as for a new QC code or MLIP.

Here we outline the core tasks in ESTEEM, as required to set up a system, train appropriate MLIPs via active learning, and then use them for spectroscopy.


\subsection{System Preparation}
\label{sec:sys_prep}
This set of tasks generates an initial dataset of energies and forces corresponding to solvent and solute-in-solvent clusters, upon which the 1$^{\mathrm{st}}$ iteration of ML models can be trained.
\paragraph{Solutes}
\label{sec:prep_solutes}
First, coordinates corresponding to initial geometries for the solute and solvent molecules are required. These are either supplied by the user, or can be downloaded from the PubChem database\cite{BOLTON2008217}. These initial geometries are then optimized, firstly in vacuum, then in each of the solvents in consideration. Both sets of optimizations are carried out on a specified adiabatic PES. The vertical excitation energies from the ground state to a specified number of chronological excited states are then calculated. All calculations in this task employ the chosen QC wrapper, and can be performed with a range of methodological choices such as basis set and exchange-correlation functional, with static spectra generated with each for comparison, to ensure good convergence and accuracy with respect to level of theory.

\paragraph{Solvate}\label{sec:solvate}
Solute geometries resulting from implicit solvent geometry optimizations (section \ref{sec:prep_solutes}) are surrounded by boxes of explicit solvent, producing one solvated system for each unique combination of solute and solvent. The optimized solvent geometries (section \ref{sec:prep_solutes}) are also used to generate bulk solvent systems of appropriate size. Typical packages for this first setup step include AMBERtools \cite{Case2023} and PackMol \cite{https://doi.org/10.1002/jcc.21224}.

An external software specified by the assigned molecular dynamics (MD) wrapper is then used to propagate four phases of ground state MD for each of the resulting systems. For the first two stages, bonds featuring hydrogen atoms have their lengths constrained to the optimized values: this helps to prevent these systems from occupying macroscopically insignificant regions of phase space. The systems are gradually heated to a target temperature in an NVT ensemble. Then the densities of the systems are equilibrated in an NPT ensemble. Following this there is a further equilibration step within the NVT ensemble, with unconstrained bond lengths. Finally the production MD is run, also in the NVT ensemble, with snapshots of the systems being taken, separated by specified time interval.

\paragraph{Clusters}\label{sec:clusters}
From the resulting snapshots (section \ref{sec:solvate}), new configurations are generated, to provide snapshots of an appropriate size and composition for evaluation with the `ground truth' electronic structure method. This is achieved through `carving': solvent molecules are deleted if none of their constituent atoms are within a specified distance ($R_\mathrm{carve}$) of any atom in the solute/central solvent molecule. Optionally, the outermost solvent molecules are then deleted until the total number of atoms falls below or equal to a specified threshold ($n_\mathrm{atoms}$), to preclude generating configurations that are too large to run efficiently on available hardware with the chosen methodology. Different sized clusters can be carved for the solution and bulk solvent systems by specifying different and $R_\mathrm{carve}^s$ values for different molecular species $s$. Singlepoint calculations of total energies, electric dipole moments and atomic forces are then carried out for the carved configurations of solution systems in the ground and first excited state. For the carved configurations of the solvent systems, only ground state singlepoint calculations are carried out, as the solvent is assumed to be photoactively inert, and to have a much wider optical gap than the target solute molecule. The chosen QC wrapper is used to run these calculations.

\subsection{Active Learning}
\label{sec:active_learning}
This series of tasks trains Machine Learning surrogate models to replace the electronic structure method entirely. Models are trained either from scratch or by fine-tuning pre-existing foundation models \cite{Radova2025,D4FD00107A} to predict energies, forces, dipoles and transition dipoles as a function of atomic positions. By using an iterative active learning procedure, we intend to reduce the quantity of data--and the associated computational expense--required to train MLIPs that are accurate across all regions of phase space relevant to the theoretical spectroscopy tasks required. The procedures outlined in this section are repeated a specified number of times as a loop as shown in the central panel of Fig.~\ref{fig:esteem_workflow}, which results in several iterations of committees of calculators.

\paragraph{ML Training}\label{sec:actlrn_MLTraining}
The user chooses a wrapper to an Machine Learned Interatomic Potential package, which will typically expose a `train' function. The parameters passed to this model define the architectures of trained MLIPs. As training data, the electronic-structure calculations from section \ref{sec:clusters} are assembled into one large dataset, comprising both solute-in-solvent configurations and solvent-in-solvent configurations for the initial dataset and those of all completed training loop iterations. The snapshots from the bulk solvent and solution trajectories are randomly allocated as training and validation snapshots. For each solution system, these training and validation datasets are used to train a specified number of independent realisations of the chosen MLIP, each typically started with initial weights determined by a different random seed. These form a committee of models, sometimes referred to as an ensemble in other works. As has become standard in MLIP training\cite{liu2025finetuninguniversalmachinelearnedinteratomic}, the first stage training epochs uses a loss function weighted more towards the accurate calculation of atomic forces than total energies. Then, the loss function is switched, and for stage two, the remaining training epochs use a loss function with greater weight placed on accuracy of total energies. The use of different loss functions in the two stages is expected to lead to a smoother and more accurate depiction of the Potential Energy Surface.

In subsequent iterations, clusters selected as training and validation data (see below) are added to the pre-existing training and validation sets for the corresponding MLIPs, and the pre-existing MLIP models are used as the starting point for further training with the additional new data. Assignment of the previous data into training and validation trajectories is retained for further steps to avoid any mixing between training and validation data.

\paragraph{ML Trajectories}\label{sec:actlrn_MLTrajectories}
The trained calculators are then used as interatomic potentials to propagate MD. The ground state calculators drive the dynamics in the ground state solvent and solution systems. The excited state MLIPs are used to simulate the solution systems in the first excited state. Each calculator within a committee starts an MD simulation (in an NVT ensemble) from a different randomly selected snapshot from the corresponding trajectory of snapshots generated in section \ref{sec:solvate}, \textit{Solvate}. Each simulation consists of an equilibration phase, followed by a production run. Snapshots of configurations are captured throughout the production run and are concatenated to make a single trajectory.
\paragraph{ML Clusters:}\label{sec:actlrn_Clusters}
The trajectories obtained in section \ref{sec:actlrn_MLTrajectories}, are used to generate new carved trajectories of bulk solvent and solution clusters using the same carving procedure outlined in section \ref{sec:clusters}, \textit{Clusters}. If multiple potentials have been trained to predict the same target quantities for the same system, the total energies of snapshots in these carved trajectories are predicted using each MLIP in the corresponding committee. The standard deviation in energies ($\sigma_\mathrm{E}$) predicted by the committee of MLIPs is calculated for each cluster. This can be used as a non-quantitative indicator of which regions of phase space the MLIPs are not representing accurately. Clusters are then selected as training and validation data through either random selection or an active learning methodology. The selected QC wrapper is used to evaluate the energies, forces and dipole moments of the selected training and validation clusters.

\paragraph{ML Testing}\label{sec:actlrn_MLTest}
 Errors associated with models are quantified through comparison of MLIP predicted energies, dipoles and forces for training and validation clusters to the QC calculated values.

\subsection{Spectra Generation}
Once sufficiently accurate MLIPs have been trained, they are used to sample macroscopically relevant regions of phase space. Excitation energies associated with the sampled configurations are then used to generate UV/Vis absorption and emission spectra. 
\paragraph{Spectra ML Trajectories}\label{sec:SG_ML_Trejactories}
The procedure outlined in \ref{sec:actlrn_MLTrajectories}, \textit{ML Trajectories}, is followed, except only one ground and excited state MLIP from each committee of calculators is used to propagate dynamics. Additionally, since the resulting trajectories will be used to predict macroscopic properties, the specified length of the production run phase is generally significantly longer.

\paragraph{Spectra Clusters}\label{sec:SG_Clusters}
The same carving procedure outlined in section \ref{sec:clusters}, \textit{Clusters}, is undertaken on the production run snapshots obtained in the previous task. However, ground and excited state energies for clusters are not predicted using a QC software; instead, MLIPs are used to infer excitation energies. The trajectories of energy gaps resulting from the ground and excited state simulations will be used to predict absorption and emission spectra, respectively.
\paragraph{Spectra}\label{sec:SG_MolSpecPy}
To generate the spectra, the `dynamic second-order cumulant expansion' (DSCE) method--a technique developed in recent years by Zuehlsdorff et al.--was utilised\cite{doi:10.1146/annurev-physchem-090419-051350,10.1063/1.5006043,10.1063/1.5114818,10.1063/5.0013739}. When provided with a trajectory of purely electronic energy gaps obtained through dynamics, DSCE attempts to incorporate vibronic features into the resulting spectra. This approach approximates the response function within the Condon approximation ($\chi(t)$) as the second order cumulant expansion of the energy gap fluctuation ($\delta U$):
\begin{equation}
    \chi (t) \approx \lvert V_{ge} \rvert ^{2}\ e^{-it\omega^{av}_{eg}}\ e^{g_{2} (t)},
\end{equation} where $V_{ge}$ corresponds to the transition dipole operator for transitions between the ground ($g$) and first excited state ($e$), $t$ is time, $\omega^{av}_{eg}$ is the thermally averaged energy gap operator and $g_{2}(t)$ is the second order cumulant of $\delta U$. 
With this approximated response function, the linear absorption spectrum ($\sigma (\omega)$) can be predicted as:
\begin{equation}
    \sigma (\omega) = \int^{\infty}_{-\infty} dt\ e^{it\omega}\ \chi(t)
\end{equation}, where $\omega$ is a wavelength of absorption/emission. 

\subsection{Types of ML Models and Training Protocols}

\paragraph*{Single State vs Energy Gap Models:}\label{sec:MLTP_SSvsEG}
ESTEEM MLIPs are not limited to the prediction of total energies, electric dipole moments and atomic forces associated with specific physical states: they can be trained to predict any quantity that varies with atomic positions. In addition to models trained to represent one specific adiabatic state, ESTEEM facilitates the generation of Energy Gap (EG) models, which predict energy gaps, changes in electric dipole moment, and differences in atomic forces between two adiabatic states. These models remove the erroneous contribution of poor error cancellation between two single state models.

\paragraph*{Single- vs Multi-head models}\label{sec:MLTP_SHvsMH}
ESTEEM permits the training of both single headed (SH) models--where each model predicts one set of properties per molecular configuration--and multiheaded (MH) models, where one MLIP simultaneously predicts multiple sets of properties for each set of atomic coordinates. MH models are trained with three heads: two represent the ground (MH-GS) and first excited adiabatic states (MH-ES1), and the final predicts differences between these states (MH-EG). Three separate SH models are required to predict these same properties; hereafter they are referred to as SH-GS, SH-ES1 and SH-EG, respectively.

\paragraph*{Iterative vs Static Training}\label{sec:MLTP_IvsS}
As described in section \ref{sec:actlrn_MLTraining}, the core ESTEEM workflow trains models in an iterative fashion; this facilitates the facile generation of informative training and validation clusters. However, once training data has been selected, whether new models are trained in an iterative procedure--being exposed to increasingly large fractions of the training data as training proceeds--or in a static fashion, with all training data immediately available, has been found to be inconsequential to the accuracy of resulting models. Thus, ESTEEM can also train ML models in a static fashion, using existing training data obtained in the iterative training of earlier models.

\section{Predicting UV-Vis Spectra: Nile Red}
\label{sec:nilered}
Nile red (NR) is a lipophilic fluorescent dye with strong solvatochromic character\cite{deye1990nile, golini1998further}. It is widely used in the study of biological matter\cite{10.1083/jcb.100.3.965,CHEN200941} and has even been used to identify and quantify microplastics\cite{HO2024109708}. Thus, there is much interest in the spectroscopic properties of solvated NR.\\
The highly solvatochromic nature of NR, combined with the fact that its absoprtion spectrum in the visible region is defined by an $S_{1}\leftarrow S_{0}$ transition, makes it an ideal candidate to evaluate the performance of ground and excited state solution-phase MLIPs. ESTEEM was used to generate three different types of MLIP: SH models trained with small clusters ($R_\mathrm{carve}$ = 2.35-3.0~\AA), as well as SH and MH models trained with the same small solution clusters ($R_\mathrm{carve}^\mathrm{solu}$ = 2.35-3.0~\AA) and larger solvent clusters ($R_\mathrm{carve}^\mathrm{solv}$ = 5.0~\AA). 
The performance of each set of models was evaluated through the prediction of UV-Vis absorption and emission spectra for NR solvated by cyclohexane, ethanol and acetonitrile. Corresponding experimental spectra were generated for comparison. 

\renewcommand{\thesubsubsection}{subsubsection}

\subsection{Computational Methodology}

\subsubsection{System Preparation}
These models were trained using the core ESTEEM workflow (section \ref{sec:overview}), with the following settings.

\paragraph*{Solutes}\label{sec:NRSP_Solutes} Coordinate files for NR, acetonitrile, ethanol and cyclohexane were downloaded from the PubChem database. The Broyden–Fletcher–Goldfarb–Shanno algorithm was used for all geometry optimizations. The QC wrapper used throughout the entirety of section \ref{sec:nilered} interfaces with ORCA/5.0.2, a gaussian basis DFT and TD-DFT package. Its implementations of DFT-PBE0 and TD-DFT-PBE0 were used in combination with the D3BJ dispersion correction for ground and excited state calculations, respectively\cite{10.1063/1.478522}. For all calculations, the def2-TZVP basis set was used\cite{C4CP04286G}. The conductor-like polarizable continuum model, parameterised with ORCA's default dielectric constant and refractive index for each solvent, was used in implicit solvent calculations.

\paragraph*{Solvate}\label{sec:NRSP_Solvate} Dynamics were propagated for six systems: NR surrounded by each solvent, and bulk solvents. The initial dimensions of the cubic simulations boxes and the number of molecules within the boxes are shown in Table \ref{table:solvation_system_sizes}. All simulations used 2~fs timesteps, a Langevin thermostat and a 9~\AA\ cutoff for long-range interactions calculated using Ewald's summation. The AMBER force field, which was interfaced with the MD wrapper, was used to propagate all stages of dynamics. In the first phase systems were heated gradually from a temperature of 0~K to $\sim$~300~K--running 100 timesteps of NVT dynamics at each temperature--over 10,000 timesteps. The second, third and fourth phases consisted of 100,000, 50,000 and 5,000,000 time steps, respectively. Snapshots were taken after every 200 timesteps of the production run, resulting in 2500 snapshots of each system.
\begin{table}[b]
\begin{center}
\setlength{\tabcolsep}{8pt}
\renewcommand{\arraystretch}{1.2}
\begin{tabular}{|c||*{2}{c|}}
\hline
 System & Box Size \AA & \multicolumn{1}{c|}{\shortstack{No. of Solvent\\ Molecules}}\\
\hline\hline
NR in ethanol & 13.0 $\times$ 13.0 $\times$ 13.0 & 320\\
\hline
Bulk ethanol & 15.2 $\times$ 15.2 $\times$ 15.2 & 233\\
\hline
NR in acetonitrile & 13.6 $\times$ 13.6 $\times$ 13.6 &380\\
\hline
Bulk acetonitrile & 13.6 $\times$ 13.6 $\times$ 13.6 & 377\\
\hline
NR in cyclohexane & 13.6 $\times$ 13.6 $\times$ 13.6 & 74 \\
\hline
Bulk cyclohexane & 15.2 $\times$ 15.2 $\times$ 15.2 & 99\\
\hline

\end{tabular}
\caption{Table showing the dimensions of simulation boxes, and the number of solvent molecules in each box. Nile red contains 42 atoms, acetonitrile contains 6 atoms, ethanol contains 9 atoms and cyclohexane contains 18 atoms.}
\label{table:solvation_system_sizes}
\end{center}
\end{table}
\paragraph*{Clusters}\label{sec:NRSP_Clusters} $n_\mathrm{atoms}$ was set to 185 and $R_\mathrm{carve}$ was set as 2.5~\AA\, 3.0~\AA\ and 2.35~\AA\ for solution systems featuring ethanol, acetonitrile and cyclohexane respectively. These settings resulted in the majority of the solute clusters having 130-180 atoms, which can be feasibly evaluated with TD-DFT calculations.

\subsubsection{Iterative Training of Small Cluster SH Models} \label{sec:Iterative_Training_Small_SH}
\paragraph*{ML Training (1$^\mathrm{st}$ iteration)}\label{sec:NRAL_MLTraining1}
For each solute-in-solvent system, five 0th iteration (MP-GNN) MLIPs were trained for the ground and first excited state. Models were first trained for 900 epochs with a loss function weighted towards the accurate prediction of forces, then for a further 100 epochs with a loss function weighted more significantly towards the accurate prediction of energies. The ML wrapper used throughout this workflow interfaces with MACE/0.2.0\cite{Batatia2022mace, Batatia2022Design}.

\paragraph*{ML Trajectories}\label{sec:NRAL_MLTrajectories} For all systems, initial equilibrium phases consisted of 1000 0.5~fs timesteps, and production phases were 5~ps in length, with 0.5~fs timesteps and snapshots being taken every 5~fs. In both phases of dynamics, a Langevin thermostat was used to keep system temperatures $\sim$300~K.

\paragraph*{ML Clusters}\label{sec:NRAL_MLClusters} The same values for $n_\mathrm{atoms}$ and $R_\mathrm{carve}$ as in section \ref{sec:NRSP_Clusters}, \textit{Clusters}, were used to generate carved training and validation trajectories. To improve the efficiency of our training methodology, we implemented an active learning procedure. Our approach was intended to balance the informativeness, diversity and representativeness of selected training and validation data, as suggested by Zaverkin et al.\cite{D2DD00034B}
$\sigma_\mathrm{E}$ dependent weights ($w$) were calculated for all configurations ($i$): $w_{i} = e^{5000\sigma_{E_{i}}}$. The `random.choice' function of Numpy/1.19.4 then re-ordered the list of configurations with a weighted random order according to $w$\cite{numpy2020}. Finally, this list was looped through in its new order, with any clusters that were generated within 10~fs (in the same simulation) of each sampled configuration being removed from the list. This was carried until 100 clusters had been selected, which were then  randomly divided into 80 training and 20 validation clusters.

\paragraph*{2$^{\mathrm{nd}}$-4$^{\mathrm{th}}$ iteration models}\label{sec:NRSP_ALTraining2} 
In total we carried out four active learning cycles, meaning the \textit{ML Training}, \textit{ML Trajectories} and \textit{ML Clusters} tasks were repeated three more times, though for these final three iterations, models only underwent 100 epochs of training using a loss function weighted towards the accurate prediction of forces, and a further 100 epochs with a loss function weighted in favour of energies. The resulting 4$^{\mathrm{th}}$ iteration models will hereafter be referred to as model set 1 (Table \ref{table:model_sets}), with the earlier iterations having no further use.
\subsubsection{Static Training of Larger Solvent Cluster SH and MH models}
The training and validation sets used to train model set 1 were augmented by recarving all bulk solvent clusters with $R_\mathrm{carve}^\mathrm{solv}$ = 5.0~\AA\ and no upper limit for $n_\mathrm{atoms}$. The ground state properties of these larger solvent clusters were calculated using the same DFT methodology as above. Solution clusters remained unchanged. Using these augmented datasets, committees of 5 SH models were trained for both adiabatic states of each solution. Training and validation datasets for training EG models were generated by subtracting ground state total energies, electric dipole moments and atomic forces from the excited state values for the same cluster. One SH-EG and MH model was trained for each system. All MLIPs were trained in a static fashion. For the first 1500 epochs, a loss function weighted more towards the accurate treatment of atomic forces was used, and for the final 100 epochs a different loss function weighted more significantly towards the accurate prediction of energies was used. The resulting SH and MH models will hereafter be referred to as model set 2 and model set 3, respectively (Table \ref{table:model_sets}).
\begin{table*}
\begin{center}
\begin{tabular}{|c||*{4}{c|}}
\hline
 & Model Set 1 & Model Set 2& Model Set 3\\
\hline\hline
\makecell{Training solute-in-solvent\\ cluster radii (\AA)}& 2.35-3.0&2.35-3.0&2.35-3.0\\
\hline
\makecell{Training solvent \\cluster radii (\AA)}& 2.35-3.0 & 5.0 &5.0\\
\hline
Committee Size & 5 & 5 & 1\\
\hline
Model type & SH & SH & MH\\
\hline
Training type & Iterative & Static & Static \\
\hline
Architecture & MACE 32x0e+32x1o+32x2e & MACE 32x0e+32x1o+32x2e &MACE 32x0e+32x1o+32x2e\\
\hline
\end{tabular}
\end{center}
\caption{Properties of model sets 1, 2 \& 3.}
\label{table:model_sets}
\end{table*}


\subsubsection{Spectra Generation}
\paragraph*{Spectra ML Trajectories}\label{sec:NRSG_ML_Trejactories}
One ground state and first excited state MLIP from each of the SH committees (model sets 1 \& 2), as well as one ground and excited state head of each MH model (model set 3), were used to propagate the MD simulations. Each simulation consisted of a 1~ps equilibration stage, and a 40~ps production stage, where snapshots were taken after every other 0.5~fs timestep. A Langevin thermostat was used to keep system temperatures $\sim$300~K.

\paragraph*{Spectra Clusters}\label{sec:NRSG_Clusters}
All production run trajectories generated in the previous step were carved with various $R_\mathrm{carve}$ values: 0.0~\AA, 5.0~\AA\ and 7.5~\AA; and 2.35~\AA, 2.5~\AA\ and 3.0~\AA\ for systems involving cyclohexane, ethanol and acetonitrile, respectively. Each 80,000 cluster trajectory was sliced to 60,000 snapshots by removing the first 20,000 snapshots, ensuring only snapshots of a fully equilibrated system were used to generate spectra. The remaining snapshots of each solution system were then split into sets of 20,000 frames, maintaining the chronological order of snapshots. The total energies of each cluster were then calculated by both the model/head used to generate the trajectories, and the corresponding analogue for the other adiabatic state. By subtracting the predicted ground state energies from the excited state energies, S$_{1}\leftarrow$S$_{0}$ excitation energies were predicted for each cluster. The SH-EG and MH-EG calculators (model sets 2 \& 3) were also used to directly infer S$_{1}\leftarrow$S$_{0}$ excitation energies for all trajectories propagated by calculators in model sets 2 \& 3, respectively. 

\paragraph*{MolSpecPy}\label{sec:SG_MSP}
The resulting 60,000 energy gaps (processed into three sets of 20,000 energy gaps) corresponding to trajectories that resulted from each ground and first excited state simulations were then used to predict absorption and emission spectra, respectively. Although the resulting lineshapes exhibited vibronic features, their peak maxima had been significantly shifted in a non-physical manner compared to vertical excitation spectra generated from the raw energy gap data. To retain these vibronic features without detriment to the accuracy of peak position predictions, we shifted the MolSpecPy spectra in energy space to have the same peak maxima position as their vertical excitation analogues.
\subsection{Results and Discussion}
 To quantify the accuracy of these models, we refer to errors in MLIP energy predictions (w.r.t (TD-)DFT data) for four sets of test clusters. Each test set consists of carved configurations that were originally selected as training and validation systems in the ML Clusters task for the fifth iterations of model training, which never occurred. Solvent test set 1 and solute-in-solvent test set 1 consisted of bulk solvent and solute-in-solvent configurations that were carved into clusters with radii of 2.35, 2.5 \& 3.0~\AA\ for clusters featuring cyclohexane, ethanol and acetonitrile, respectively. Solvent test set 2 and solute-in-solvent test set 2 consisted of larger systems, with clusters in the former being carved to have 5~\AA\ radii, and those in the latter being carved to have an average of 220-230 atoms: 2.5, 3.0 and 4.5~\AA\ radii for NR in cyclohexane, ethanol and acetonitrile, respectively.\\
 If ground-truth ground and excited state potentials were used, as the radii of clusters used to generate spectra increase, convergence of the distribution of energy gaps would be observed. As shown in figure \ref{fig:nr_etoh_models_radii}, the shape and peak positions of spectra predicted for NR in ethanol using model set 1 diverge as cluster radii increase past 2.5~\AA, indicating a source of error that grows with the size of the cluster. Model set 1 calculators were only exposed to training data featuring solvent-solvent interactions spanning a maximum distance of $\sim$6~\AA. Thus, the treatment of longer range solvent-solvent interactions (up to $\sim$15~\AA) in larger clusters by model set 1 is not only inaccurate, but also likely inconsistent between calculators for different adiabatic states. This means there is no guarantee of favourable error cancellation when subtracting SH-GS energies from SH-ES1 energies.\\
For all three solvents, errors in energy predictions for both adiabatic states of clusters in solvent test set 2--which are physically degenerate due to the absence of solute--exceeded 0.23 eV (Tables \ref{table:MAE_Solv}). This error resulted in significantly worse predictions of energy gaps for solvent clusters featuring ethanol and acetonitrile than with model set 2. With radii of 4.5~\AA, the NR in acetonitrile test clusters feature many long range solvent-solvent interactions, and so offer the most insight into the potential improvement available with model set 2 compared to model set 1. As shown in Table \ref{table:MAE_SIS}, errors in ground and first excited state evaluations of clusters in solute-in-solvent test set 2 reduced by over 34\%, and error in energy gap predictions fell by $\sim$89\%. The energy predictions for both adiabatic states of NR in ethanol and cyclohexane also improved significantly using model set 2, and so did the prediction of energy gaps (by subtracting GS energies from ES1 energies) for NR in ethanol. It is notable that model set 2 also generally made higher accuracy predictions of state energies than model set 1 for solute-in-solvent test set 1, where long-range solvent-solvent interactions were comparativelty minimal. \\
The improvement in accuracy by using model set 2 instead of model set 1 is also evidenced in Figure \ref{fig:committees_over_models}, showing absorption and emission vertical excitation spectra generated from 5~\AA\ clusters using each MLIP in the committees. Closer agreement between spectra independently predicted by MLIPs within committees qualitatively indicates improved accuracy in models: if separate MLIPs predict more similar results over ensembles of configurations, this suggests (but does not confirm) that sources of error have been removed/reduced. Additionally, the spread of spectra predicted with calculators from the same committees shown in Figures \ref{fig:committees_over_models} and \ref{fig:committees_over_radii} supports our decision to predict energy gaps using only one model from each committee, rather than by concatenating energy gap predictions made by each member of a committee. Provided the spread of spectra between calculators is not caused by incomplete sampling of phase space, combining the energy gap data of multiple calculators will only result in the erroneous broadening and loss of vibronic resolution in resulting spectra.\\ 
The calculation of energy gaps by taking the difference in energies predicted by two separate MLIPs/heads is inherently flawed--this method unavoidably introduces error through unpredictable error cancellation. This motivated the generation of the EG models: by training a single calculator to directly infer energy gaps from cluster geometries, the influence of poor error cancellation between models is nullified. The SH-EG potentials were extremely effective, significantly reducing the mean error in all systems in all four test sets, other than for NR in acetonitrile solution test set 2, where favourable error cancellation between SH-GS and SH-ES1 models is apparent (Tables \ref{table:MAE_Solv}, \ref{table:MAE_SIS}, \ref{table:MAE_Solv_small} \& \ref{table:MAE_SIS_small}). This improvement is further evidenced in Figures \ref{fig:committees_over_models} \& \ref{fig:committees_over_models_SD}, with a denser clustering of the predicted absorption and emission spectra generated with model set 2 SH-EG committee than with the model set 2 SH-ES1 and SH-GS committees. Additionally, as shown in Figure \ref{fig:spectra_solvents_radii}, the shape and peak positions of spectra generated using energy gaps calculated with SH-EG calculators converge with increasing radii. Thus, for the purpose of accurately predicting energy gaps, the use of specifically trained EG calculators--as an alternative to subtracting the SH-GS values from SH-ES1 energies--is optimal.\\
\begin{figure}
    \centering
    \includegraphics[width=0.49\textwidth]{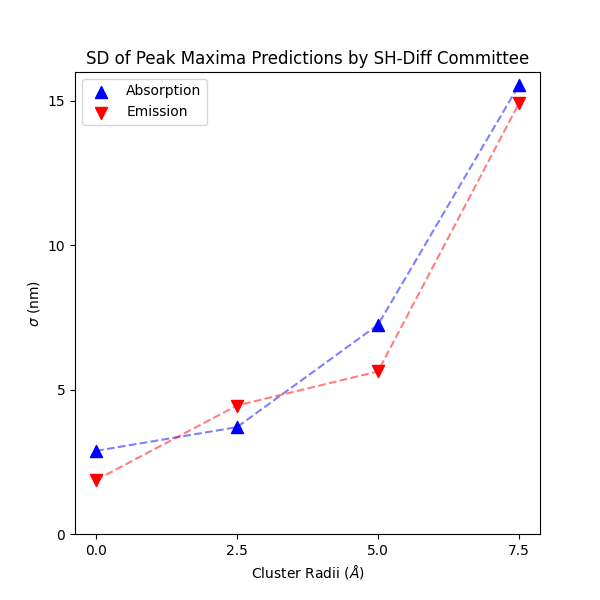}
    \caption{Sample standard deviations in spectra peak maxima predicted using each of the 5 SH-EG models in the committees in model set 2 to infer energy gaps for NR in ethanol clusters carved with radii of 0.0, 2.5, 5.0 and 7.5~\AA.}
    \label{fig:stdev_of_peaks}
\end{figure}
Figure \ref{fig:nr_etoh_models_radii} c.) \& d.) show a comparison of spectra for NR in ethanol generated using energy gaps predicted by model set 2's SH-EG calculator and model set 3's MH-EG head, respectively. While the spectra generated using the SH-EG model display no observable divergence up to a cluster radii of 7.5~\AA, those generated with the MH-EG head show a loss of vibronic resolution and non-negligible divergence of absorption and emission peak positions with increasing cluster radii. Further study of the stability of predictions made by these models with increasing cluster radii suggests that error in model set 3 grows more rapidly with increasing cluster radius than model set 2. Figure \ref{fig:sources_of_error_MH}a.) shows that for NR in ethanol the disagreement in energy gaps predicted by subtraction of energies from separate models/heads and those directly inferred with a EG model reached 0.177~eV at 7.5~\AA\ for model set 3, and only 0.028~eV at 7.5~\AA\ for model set 2. As demonstrated in Figure \ref{fig:sources_of_error_MH} b.), when predicting the energy gaps associated with augmented NR in ethanol clusters--where the NR molecule had been deleted--the error in prediction with a MH-EG model grew exponentially with cluster radii, whereas with SH-EG it only grew linearly. Additionally, for NR in ethanol and acetonitrile, Figure \ref{fig:error_models_radius_systems} shows that EG models from model set 2 predict mean energy gaps that converge with cluster radii up to 7.5~\AA, whereas predictions from MH-EG diverge to some extent. The training of model sets 2 \& 3 was identical apart from the fact that they are SH and MH models. By using the same architecture (32x0e+32x1o+32x2e) to train MH models with three heads as was used to train SH models with only one target state, it is plausible that the performance of the MH models are limited by the size and flexibility of the underlying architecture. We briefly investigated this by training a new MH model for NR in ethanol, following the same procedure as above, but with a 64x0e+64x1o+64x2e architecture. As shown in Tables \ref{table:MAE_Solv} \& \ref{table:MAE_SIS}, this larger model offered significant improvements in the prediction of energies of both adiabatic states in solvent and solute-in-solvent clusters. This suggests that the use of a larger architecture, offering enhanced model flexibility, could improve the stability of MH model predictions with increasing cluster radii and reduce the divergence of energy gap predictions. Though this larger architecture did not offer improved accuracy for solute-in-solvent energy gap calculations, this results from the test clusters being 3.0~\AA\ in radius: the identified metrics of error in the MH-EG model (figure 5) and divergence in predicted energy gaps (figure S6) only increase appreciably at radii greater than 5.0~\AA.\\
With model set 2 (using SH-EG) offering the most stable predictions of absorption and emission spectra with increasing cluster radii (Figures \ref{fig:nr_etoh_models_radii} and \ref{fig:spectra_solvents_radii}), this is the most effective model set we generated. In order to determine which radii to carve clusters at for energy gap evaluations, we aimed to generate highly converged spectra while keeping radii-linked errors negligible. Figures S5 \& S8 show that convergence in peak positions and shape is achieved by 5~\AA\ for all systems. Vertical excitation spectra for NR in ethanol--generated with clusters of various sizes--using each set of SH-GS, SH-ES1 (both for propagating dynamics) and SH-EG (to predict energy gaps) MLIPs from model set 2, are displayed in Figure \ref{fig:committees_over_radii}. The sample standard deviation of predicted peak maxima across the committee is shown in Figure \ref{fig:stdev_of_peaks}; a roughly linear trend is observed up to a radius of 5~\AA\ for both absorption and emission spectra, reaching $\pm$7.2 and $\pm$5.6~nm, respectively. Between 5~\AA\ and 7.5~\AA\ standard deviation grows significantly to $\pm$15.5 and $\pm$14.9~nm for the absorption and emission spectra, respectively. Thus, $R_{\mathrm{carve}}$ values of $\sim$5.0~\AA\ appear to produce models with an optimal combination of converged spectra (wrt radii) and low error. 
Figure \ref{fig:final_spectra} displays both experimental spectra and spectra generated using model set 2, with an SH-EG used calculator to directly infer energy gaps from clusters carved to 5.0~\AA\ radii. While strict scrutiny of the resemblance of predicted spectra to experimental spectra is not practical since the MLIPs are trained to reproduce the (TD-)DFT predicted PESs, which differ slightly from ground truth PESs, the degree to which the predicted spectra reproduce trends between systems is informative. ESTEEM correctly predicted the pattern of Stokes shifts between the three systems, with the magnitude of shifts increasing from NR in cyclohexane to ethanol, and from NR in ethanol to acetonitrile (Figure \ref{fig:stokesshifts_systems}). While ESTEEM did not reproduce the magnitude of experimental solvatochromic shifts, it correctly predicted that absorption and emission spectra for NR in cyclohexane would be significantly blueshifted by similar numbers of wavelengths with respect to the other two systems (Figure \ref{fig:solvatochromicshifts_systems}). MolSpecPy proved to be relatively effective at predicting vibronic transitions from our energy gap predictions (corresponding to purely electronic transitions); it consistently extended absorptions and emissions further into the blue and red regions, respectively, which appears to correspond with physical shoulders/peaks in experimental data.\\







\begin{figure*}[]
    \centering
    \subfigure[]{\includegraphics[width=0.49\textwidth]{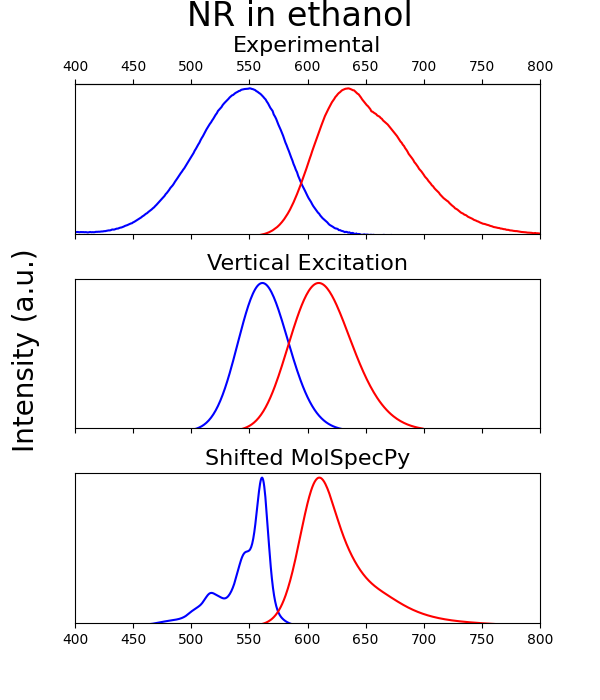}}
    \subfigure[]{\includegraphics[width=0.49\textwidth]{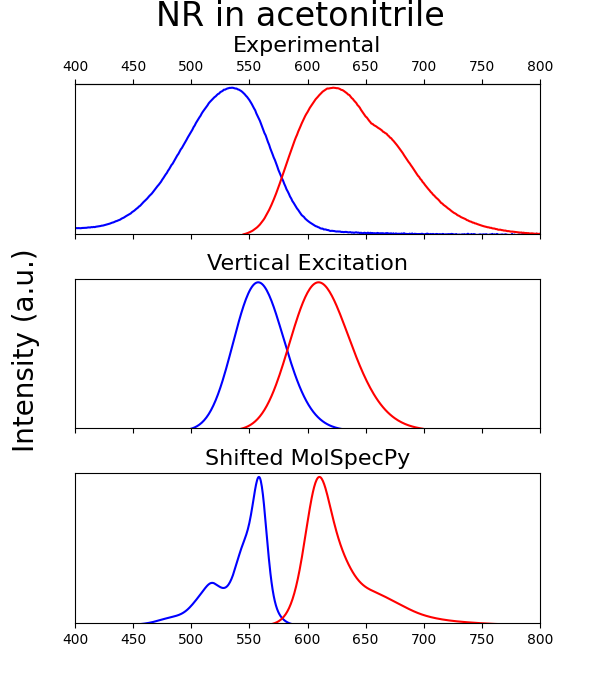}}
    \subfigure[]{\includegraphics[width=0.49\textwidth]{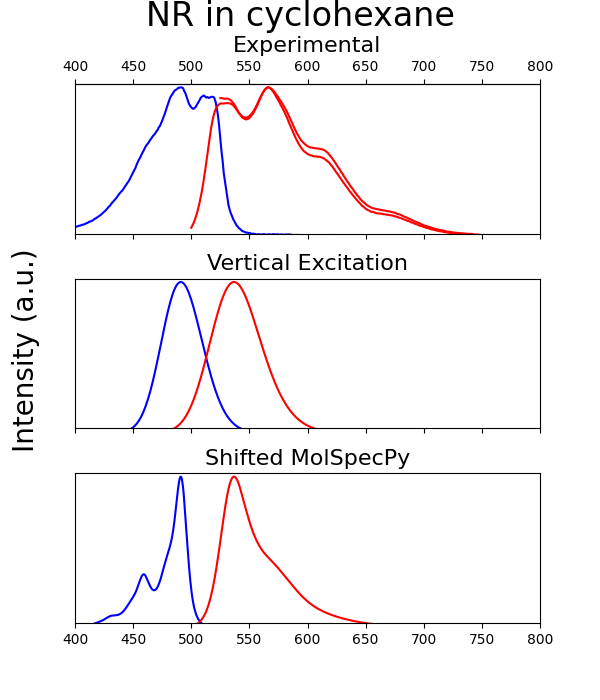}}
    \caption{Absorption (blue) and emission (red) spectra generated for a.) NR in ethanol, b.) NR in acetonitrile and c.) NR in cyclohexane. The top spectra in each subplot were generated experimentally. The middle spectra in each subplot were obtained by plotting vertical excitations predicted using model set 2, with an SH-EG used calculator to directly infer energy gaps. The bottom spectra in each subplot were generated by passing those same energy gaps into MolSpecPy, then shifting the resulting plots in energy space to have the same peak maxima as the middle spectra.}

    \label{fig:final_spectra}
\end{figure*}
\FloatBarrier
\section{Conclusion}
We have successfully designed and implemented a workflow that permits the feasible prediction of energies and energy gaps of/between adiabatic states--and consequently UV/Vis absorption and fluorescence spectra--with quasi-quantum accuracy. Despite all solute-in-solvent training data being carved with $R^\mathrm{solu}_\mathrm{carve}$=3.0~\AA\ and $n_\mathrm{atoms}$=185, our model set 2 MLIPs are able to consistently make stable and accurate adiabatic energy and energy gap predictions for clusters carved with $R\mathrm{carve}$=5.0~\AA, consisting of $\sim$220-500 atoms. With the accuracy of these models non-neligibly waning between $R_\mathrm{carve}$ values of 5.0 and 7.5~\AA (Figure \ref{fig:stdev_of_peaks}), our MLIPs are not suitable for simulating immense periodic systems; however, as shown in Figures \ref{fig:spectra_solvents_radii} \& \ref{fig:peakmaxima_radii_systems}, predicted spectral lineshapes of all three systems are converged for clusters carved with $R_\mathrm{carve}$=5.0~\AA. Hence, the ESTEEM workflow is capable of training sufficiently stable and accurate MLIPs to predict converged and accurate UV/Vis absorption and emission spectra.


\section{Acknowledgements}

J. E. acknowledges funding from the EPSRC CDT in Modelling of Heterogeneous Systems funded by EP/S022848/1. Computing facilities were provided by the Scientific Computing Research Technology Platform of the University of Warwick using the High Performance Computing (HPC) cluster Avon, and the Sulis Tier 2 platforms at HPC Midlands+ funded by the Engineering and Physical Sciences Research Council (EPSRC), grant number EP/T022108/1.



\bibliography{esteem}

\newpage
\newpage
\newpage

\title[Supplementary Information]
  {Supplementary Information}
\maketitle
\setcounter{figure}{0}
\renewcommand{\figurename}{Fig.}
\renewcommand{\thefigure}{S\arabic{figure}}
\setcounter{table}{0}
\renewcommand{\tablename}{Table}
\renewcommand{\thetable}{S\arabic{table}}

\begin{table*}
    \centering
    \begin{subtable}
    \linebreak
    \centering
    \setlength{\tabcolsep}{8pt}
    \renewcommand{\arraystretch}{1.2}
    \begin{tabular}{|c||*{4}{c|}}
    \hline
    System (GS)& Model Set 1 & Model Set 2 & Model Set 3 & MACE MH 64 \\
    \hline\hline
    Ethanol & 343 & 36 & 57& 52\\
    \hline
    Acetonitrile & 277 & 42 & 35& -\\
    \hline
    Cyclohexane & 497& 42 & 523& -\\
    \hline
    \end{tabular}
    \end{subtable}
    \vspace{0.5cm}
    \begin{subtable}
    \linebreak
    \linebreak
    \\
    \centering
    \setlength{\tabcolsep}{8pt}
    \renewcommand{\arraystretch}{1.2}
    \begin{tabular}{|c||*{4}{c|}}
    \hline
    System (ES1) & Model Set 1 & Model Set 2 & Model Set 3 & MACE MH 64 \\
    \hline\hline
    Ethanol & 239& 35& 79& 30\\
    \hline
    Acetonitrile & 241 & 36& 32& -\\
    \hline
    Cyclohexane &462 & 32 &300 & -\\
    \hline
    \end{tabular}
    \end{subtable}
    \vspace{0.5cm}
    \begin{subtable}
    \linebreak
    \linebreak
    \\
    \centering
    \setlength{\tabcolsep}{8pt}
    \renewcommand{\arraystretch}{1.2}
    \begin{tabular}{|c||*{5}{c|}}
    \hline
    System ($\Delta E$)& \makecell{Model Set 1\\(ES1-GS)} & \makecell{Model Set 2\\(ES1-GS)} & \makecell{Model Set 2\\(EG)} & \makecell{Model Set 3\\(EG)}& \makecell{MACE MH 64\\(EG)}\\
    \hline\hline
    Ethanol & 10.4& 8.0& 4.6& 7.4&8.2\\
    \hline
    Acetonitrile & 36.6& 16.4& 6.0& 8.0&-\\
    \hline
    Cyclohexane & 34.2& 36.6& 2.7& 1.4&-\\
    \hline
    \end{tabular}
    \end{subtable}
    \caption{MAE (meV) in MLIP energy predictions of 5.0 \AA\ solvent clusters carved from configurations selected as training and validation data for the fifth iteration of model training that never occured.}
\label{table:MAE_Solv_mev}
\end{table*}

\begin{table*}
    \centering
    \begin{subtable}
    \linebreak
    \centering
    \setlength{\tabcolsep}{8pt}
    \renewcommand{\arraystretch}{1.2}
    \begin{tabular}{|c||*{4}{c|}}
    \hline
    System (GS)& Model Set 1 & Model Set 2 & Model Set 3 & MACE MH 64 \\
    \hline\hline
    NR in ethanol & 483& 313& 360& 197\\
    \hline
    NR in acetonitrile & 643 & 363 & 431 & -\\
    \hline
    NR in cyclohexane & 295& 55 & 368& -\\
    \hline
    \end{tabular}
    \end{subtable}
    \vspace{0.5cm}
    \begin{subtable}
    \linebreak
    \linebreak
    \\
    \centering
    \setlength{\tabcolsep}{8pt}
    \renewcommand{\arraystretch}{1.2}
    \begin{tabular}{|c||*{4}{c|}}
    \hline
    System (ES1) & Model Set 1 & Model Set 2 & Model Set 3 & MACE MH 64 \\
    \hline\hline
    NR in ethanol & 400& 342& 405& 225\\
    \hline
    NR in acetonitrile & 552 & 363& 379& -\\
    \hline
    NR in cyclohexane &256& 84& 142& -\\
    \hline
    \end{tabular}
    \end{subtable}
    \vspace{0.5cm}
    \begin{subtable}
    \linebreak
    \linebreak
    \\
    \centering
    \setlength{\tabcolsep}{8pt}
    \renewcommand{\arraystretch}{1.2}
    \begin{tabular}{|c||*{5}{c|}}
    \hline
    System ($\Delta E$) & \makecell{Model Set 1\\(ES1-GS)} & \makecell{Model Set 2\\(Es1-GS)} & \makecell{Model Set 2\\(EG)} & \makecell{Model Set 3\\(EG)}& \makecell{MACE MH 64\\(EG)}\\
    \hline\hline
    NR in ethanol & 83.2& 30.2& 10.9& 7.8&9.7\\
    \hline
    NR in acetonitrile & 91.2& 10.0& 10.5& 20.8&-\\
    \hline
    NR in cyclohexane & 38.9&42.3 & 6.4& 10.3&-\\
    \hline
    \end{tabular}
    S\end{subtable}
    \caption{MAE (meV)in MLIP energy predictions of solute-in-solventclusters carved from configurations selected as training and validation data for the fifth iteration of model training that never occured. For systems featuring cyclohexane, ethanol and acetonitrile, configurations were carved with radii of 2.5, 3.0 and 4.5~\AA, respectively.}
\label{table:MAE_SIS_mev}
\end{table*}

\begin{table*}
    \centering
    \begin{subtable}
    \linebreak
    \centering
    \setlength{\tabcolsep}{8pt}
    \renewcommand{\arraystretch}{1.2}
    \begin{tabular}{|c||*{4}{c|}}
    \hline
    System (GS)& Model Set 1 & Model Set 2 & Model Set 3 & MACE MH 64 \\
    \hline\hline
    Ethanol & 0.34329 & 0.03555& 0.05745& 0.05260\\
    \hline
    Acetonitrile & 0.27723 & 0.04160 & 0.03498& -\\
    \hline
    Cyclohexane & 0.49662& 0.04204 & 0.52264& -\\
    \hline
    \end{tabular}
    \end{subtable}
    \vspace{0.5cm}
    \begin{subtable}
    \linebreak
    \linebreak
    \\
    \centering
    \setlength{\tabcolsep}{8pt}
    \renewcommand{\arraystretch}{1.2}
    \begin{tabular}{|c||*{4}{c|}}
    \hline
    System (ES1) & Model Set 1 & Model Set 2 & Model Set 3 & MACE MH 64 \\
    \hline\hline
    Ethanol & 0.23882& 0.03463& 0.07905& 0.03037\\
    \hline
    Acetonitrile & 0.24063 & 0.03629& 0.03244& -\\
    \hline
    Cyclohexane &0.46240 & 0.03174 &0.29967 & -\\
    \hline
    \end{tabular}
    \end{subtable}
    \vspace{0.5cm}
    \begin{subtable}
    \linebreak
    \linebreak
    \\
    \centering
    \setlength{\tabcolsep}{8pt}
    \renewcommand{\arraystretch}{1.2}
    \begin{tabular}{|c||*{5}{c|}}
    \hline
    System ($\Delta E$)& \makecell{Model Set 1\\(ES1-GS)} & \makecell{Model Set 2\\(ES1-GS)} & \makecell{Model Set 2\\(EG)} & \makecell{Model Set 3\\(EG)}& \makecell{MACE MH 64\\(EG)}\\
    \hline\hline
    Ethanol & 0.10447& 0.00804& 0.00456& 0.00741&0.00816\\
    \hline
    Acetonitrile & 0.03660& 0.01639& 0.00596& 0.00803&-\\
    \hline
    Cyclohexane & 0.03422& 0.03664& 0.00267& 0.00143&-\\
    \hline
    \end{tabular}
    \end{subtable}
    \caption{MAE (eV) in MLIP energy predictions of 5.0 \AA\ solvent clusters carved from configurations selected as training and validation data for the fifth iteration of model training that never occured.}
\label{table:MAE_Solv}
\end{table*}

\begin{table*}
    \centering
    \begin{subtable}
    \linebreak
    \centering
    \setlength{\tabcolsep}{8pt}
    \renewcommand{\arraystretch}{1.2}
    \begin{tabular}{|c||*{4}{c|}}
    \hline
    System (GS)& Model Set 1 & Model Set 2 & Model Set 3 & MACE MH 64 \\
    \hline\hline
    NR in ethanol & 0.48338& 0.31264& 0.35963& 0.19693\\
    \hline
    NR in acetonitrile & 0.64335 & 0.36250 & 0.43053 & -\\
    \hline
    NR in cyclohexane & 0.29540& 0.05522 & 0.36761& -\\
    \hline
    \end{tabular}
    \end{subtable}
    \vspace{0.5cm}
    \begin{subtable}
    \linebreak
    \linebreak
    \\
    \centering
    \setlength{\tabcolsep}{8pt}
    \renewcommand{\arraystretch}{1.2}
    \begin{tabular}{|c||*{4}{c|}}
    \hline
    System (ES1) & Model Set 1 & Model Set 2 & Model Set 3 & MACE MH 64 \\
    \hline\hline
    NR in ethanol & 0.40017& 0.34178& 0.40537& 0.22466\\
    \hline
    NR in acetonitrile & 0.55212 & 0.36340& 0.37944& -\\
    \hline
    NR in cyclohexane &0.25647& 0.08370& 0.14183& -\\
    \hline
    \end{tabular}
    \end{subtable}
    \vspace{0.5cm}
    \begin{subtable}
    \linebreak
    \linebreak
    \\
    \centering
    \setlength{\tabcolsep}{8pt}
    \renewcommand{\arraystretch}{1.2}
    \begin{tabular}{|c||*{5}{c|}}
    \hline
    System ($\Delta E$) & \makecell{Model Set 1\\(ES1-GS)} & \makecell{Model Set 2\\(ES1-GS)} & \makecell{Model Set 2\\(EG)} & \makecell{Model Set 3\\(EG)}& \makecell{MACE MH 64\\(EG)}\\
    \hline\hline
    NR in ethanol & 0.08321& 0.03017& 0.01092& 0.00777&0.00967\\
    \hline
    NR in acetonitrile & 0.09123& 0.01004& 0.01049& 0.020758&-\\
    \hline
    NR in cyclohexane & 0.03893&0.04230 & 0.00642& 0.01033&-\\
    \hline
    \end{tabular}
    S\end{subtable}
    \caption{MAE (eV)in MLIP energy predictions of solute-in-solventclusters carved from configurations selected as training and validation data for the fifth iteration of model training that never occured. For systems featuring cyclohexane, ethanol and acetonitrile, configurations were carved with radii of 2.5, 3.0 and 4.5~\AA, respectively.}
\label{table:MAE_SIS}
\end{table*}

\begin{table*}
    \centering
    \begin{subtable}
    \linebreak
    \centering
    \setlength{\tabcolsep}{8pt}
    \renewcommand{\arraystretch}{1.2}
    \begin{tabular}{|c||*{4}{c|}}
    \hline
    System (GS)& Model Set 1 & Model Set 2 & Model Set 3 & MACE MH 64 \\
    \hline\hline
    Ethanol & 0.02620 &0.12636 & 0.10343& 0.11901\\
    \hline
    Acetonitrile &  0.02344&  0.09864&0.07829 & -\\
    \hline
    Cyclohexane & 0.03081& 0.16645&0.32962 & -\\
    \hline
    \end{tabular}
    \end{subtable}
    \vspace{0.5cm}
    \begin{subtable}
    \linebreak
    \linebreak
    \\
    \centering
    \setlength{\tabcolsep}{8pt}
    \renewcommand{\arraystretch}{1.2}
    \begin{tabular}{|c||*{4}{c|}}
    \hline
    System (ES1)& Model Set 1 & Model Set 2 & Model Set 3 & MACE MH 64 \\
    \hline\hline
    Ethanol & 0.03115& 0.12757& 0.09579& 0.11135\\
    \hline
    Acetonitrile & 0.01828 & 0.09310& 0.08782& -\\
    \hline
    Cyclohexane & 0.03041& 0.17285& 0.25868& -\\
    \hline
    \end{tabular}
    \end{subtable}
    \vspace{0.5cm}
    \begin{subtable}
    \linebreak
    \linebreak
    \\
    \centering
    \setlength{\tabcolsep}{8pt}
    \renewcommand{\arraystretch}{1.2}
    \begin{tabular}{|c||*{5}{c|}}
    \hline
    System ($\Delta E$)& \makecell{Model Set 1\\(ES1-GS)} & \makecell{Model Set 2\\(ES1-GS)} & \makecell{Model Set 2\\(EG)} & \makecell{Model Set 3\\(EG)}& \makecell{MACE MH 64\\(EG)}\\
    \hline\hline
    Ethanol & 0.02064& 0.00301& 0.00189& 0.00118&0.00329\\
    \hline
    Acetonitrile &0.01542& 0.00553& 0.00235& 0.00257&-\\
    \hline
    Cyclohexane &0.00505 &0.00642 &0.00167 &0.00248 &-\\
    \hline
    \end{tabular}
    \end{subtable}
    \caption{MAE (eV) in MLIP energy predictions of solvent clusters carved from configurations selected as training and validation data for the fifth iteration of model training that never occured. For systems featuring cyclohexane, ethanol and acetonitrile, configurations were carved with radii of 2.35, 2.5 and 3.0~\AA, respectively.}
\label{table:MAE_Solv_small}
\end{table*}

\begin{table*}
    \centering
    \begin{subtable}
    \linebreak
    \centering
    \setlength{\tabcolsep}{8pt}
    \renewcommand{\arraystretch}{1.2}
    \begin{tabular}{|c||*{4}{c|}}
    \hline
    System (GS)& Model Set 1 & Model Set 2 & Model Set 3 & MACE MH 64 \\
    \hline\hline
    NR in ethanol & 0.07243& 0.04475& 0.06853& 0.03579\\
    \hline
    NR in acetonitrile &  0.08464&  0.05868&0.06414  & -\\
    \hline
    NR in cyclohexane & 0.08142&  0.04042& 0.25480& -\\
    \hline
    \end{tabular}
    \end{subtable}
    \vspace{0.5cm}
    \begin{subtable}
    \linebreak
    \linebreak
    \\
    \centering
    \setlength{\tabcolsep}{8pt}
    \renewcommand{\arraystretch}{1.2}
    \begin{tabular}{|c||*{4}{c|}}
    \hline
    System (ES1)& Model Set 1 & Model Set 2 & Model Set 3 & MACE MH 64 \\
    \hline\hline
    NR in ethanol &0.05617 &0.05726 &0.08995 &0.03429\\
    \hline
    NR in acetonitrile & 0.10019& 0.06356&0.05552 & -\\
    \hline
    NR in cyclohexane & 0.07594& 0.03494& 0.11596& -\\
    \hline
    \end{tabular}
    \end{subtable}
    \vspace{0.5cm}
    \begin{subtable}
    \linebreak
    \linebreak
    \\
    \centering
    \setlength{\tabcolsep}{8pt}
    \renewcommand{\arraystretch}{1.2}
    \begin{tabular}{|c||*{5}{c|}}
    \hline
    System ($\Delta E$)& \makecell{Model Set 1\\(ES1-GS)} & \makecell{Model Set 2\\(ES1-GS)} & \makecell{Model Set 2\\(EG)} & \makecell{Model Set 3\\(EG)}& \makecell{MACE MH 64\\(EG)}\\
    \hline\hline
    NR in ethanol & 0.02790& 0.02722& 0.00916& 0.01008& 0.01001\\
    \hline
    NR in acetonitrile &0.06534 &0.01810 & 0.00817& 0.00953&-\\
    \hline
    NR in cyclohexane & 0.01821& 0.03080& 0.00629& 0.00732&-\\
    \hline
    \end{tabular}
    \end{subtable}
    \caption{MAE (eV)in MLIP energy predictions of solute-in-solvent clusters carved from configurations selected as training and validation data for the fifth iteration of model training that never occured. For systems featuring cyclohexane, ethanol and acetonitrile, configurations were carved with radii of 2.35, 2.5 and 3.0~\AA, respectively.}
\label{table:MAE_SIS_small}
\end{table*}

\begin{figure*}
    \centering
    \subfigure[]{\includegraphics[width=0.49\textwidth]{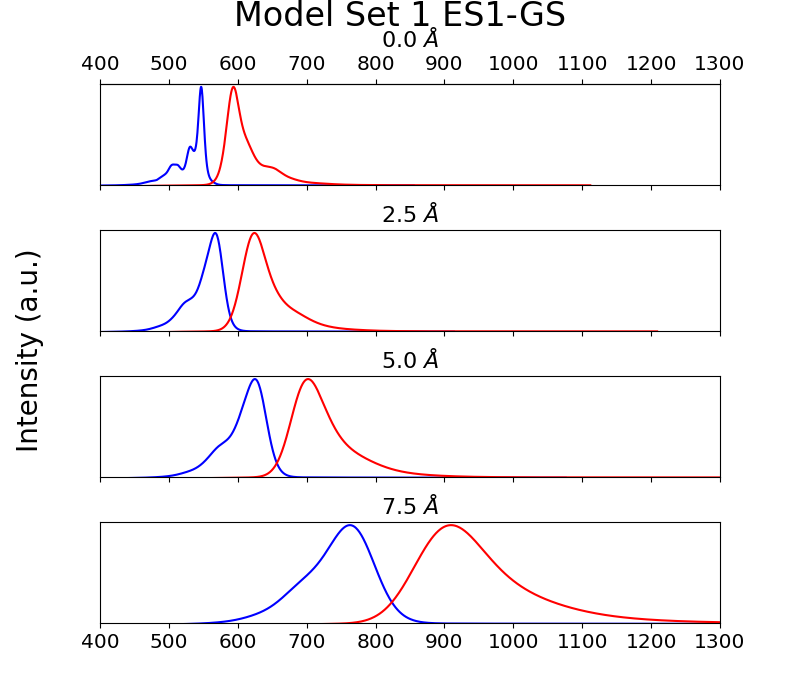}}
    \subfigure[]{\includegraphics[width=0.49\textwidth]{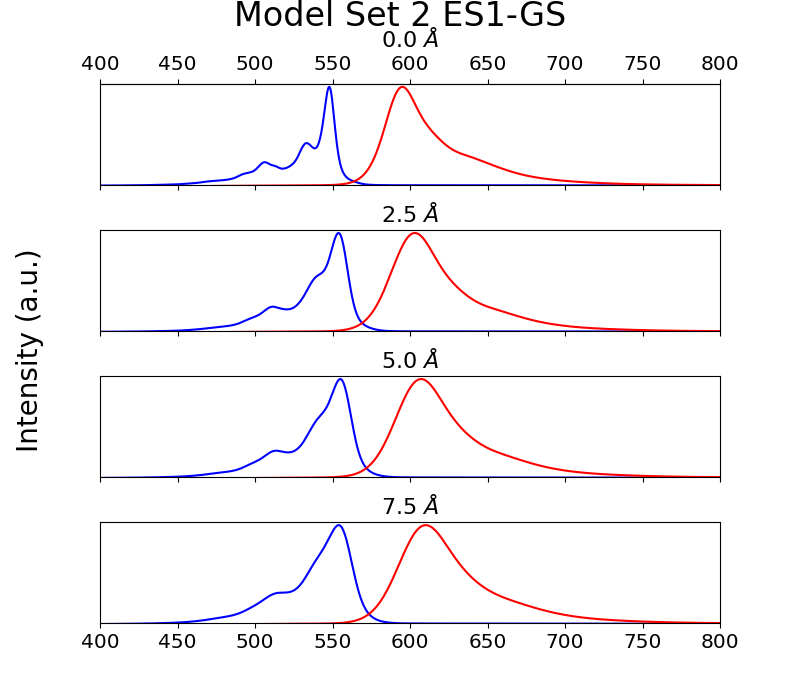}}
    \subfigure[]{\includegraphics[width=0.49\textwidth]{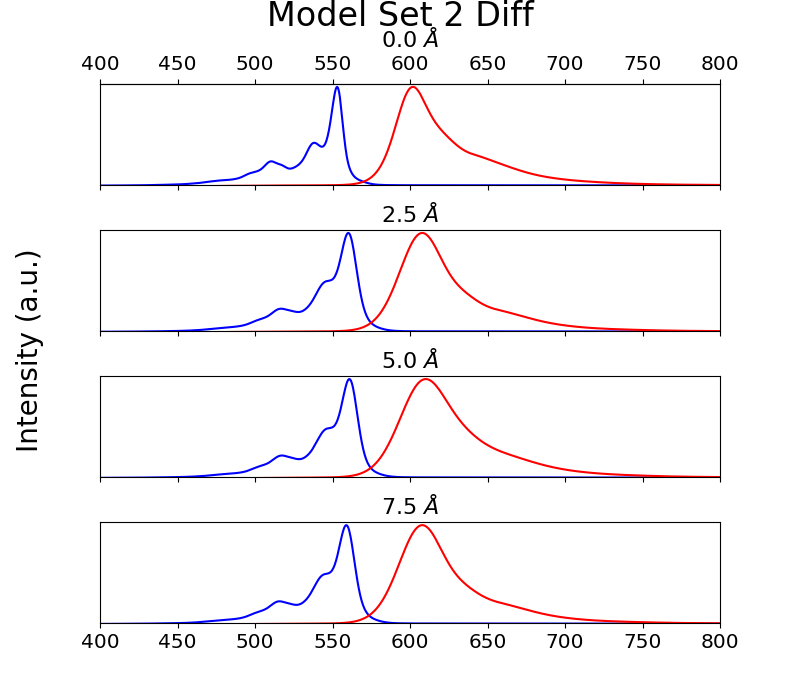}}
    \subfigure[]{\includegraphics[width=0.49\textwidth]{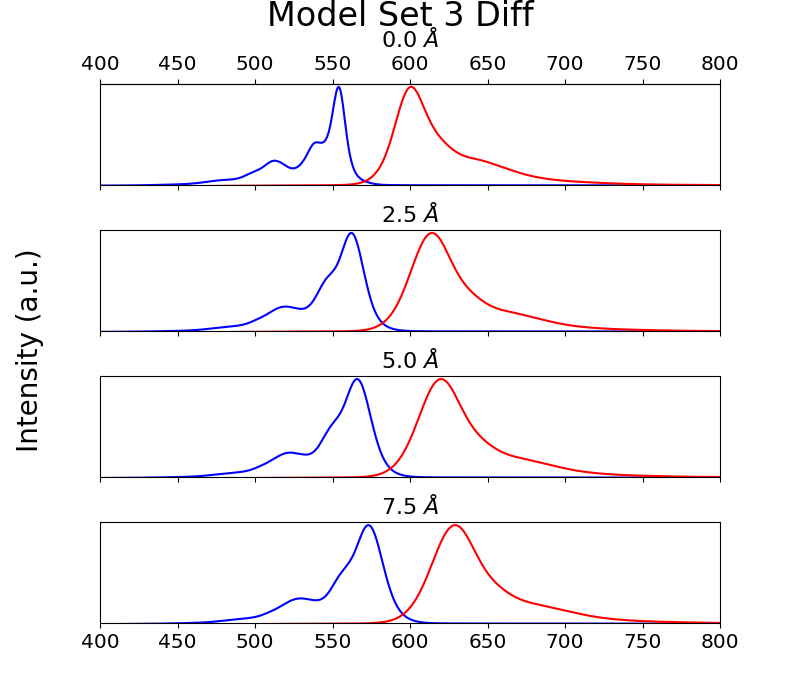}}
    \caption{Absorption and emission spectra generated for nile red in ethanol by evaluating solute-in-solvent clusters with various radii using four different methods: (a) model set 1, subtracting SH-GS from SH-ES1 energies; (b) model set 2, subtracting SH-GS energies from SH-ES1 energies; (c) model set 2, evaluating clusters with the SH-EG model; and (d) model set 3, evaluating clusters with the MH-EG head.}
    \label{fig:nr_etoh_models_radii}
\end{figure*}

\begin{figure*}
    \centering
    \subfigure[]{\includegraphics[width=0.99\textwidth]{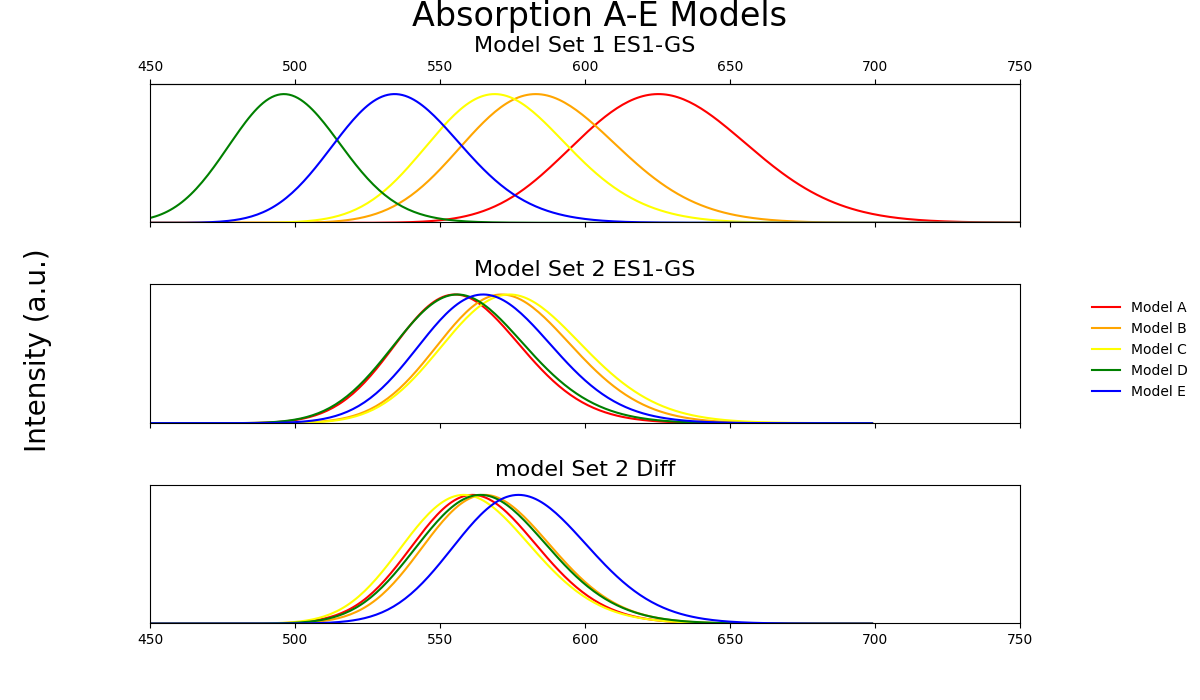}}
    \subfigure[]{\includegraphics[width=0.99\textwidth]{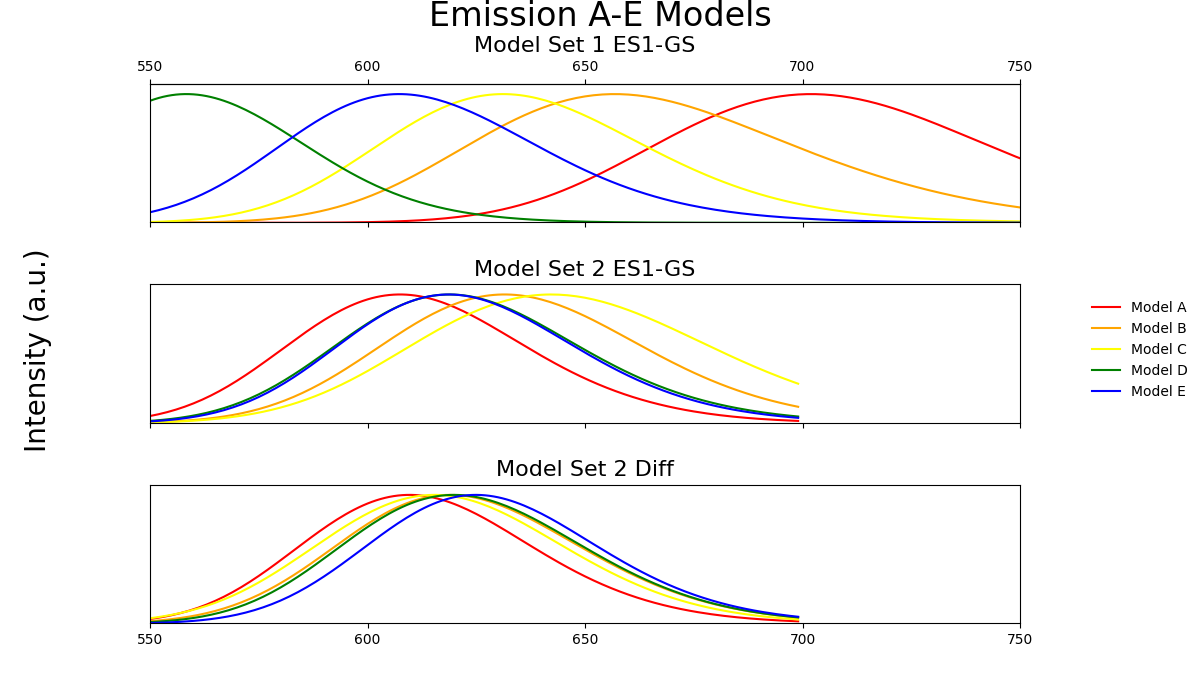}}
    \caption{Vertical excitation spectra predicted for 5.0~\AA\ radii NR in ethanol clusters using each of the 5 committee members (A-E) of model sets 1 and 2 with the method of subtracting SH-GS from SH-ES1, and also using model set 2 with the SH-EG models to directly infer energy gaps.}
    \label{fig:committees_over_models}
\end{figure*}
\begin{figure*}
    \centering
    \includegraphics[]{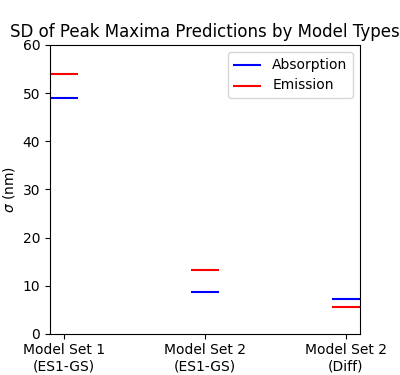}
    \caption{Sample standard deviations in spectra peak maxima positions predicted for 5.0~\AA\ radii NR in ethanol clusters using each of the 5 committee members (A-E) of model sets 1 and 2 by subtracting SH-GS from SH-ES1, and also using model set 2's SH-EG models to directly infer energy gaps.}
    \label{fig:committees_over_models_SD}
\end{figure*}

\begin{figure*}
    \centering
    \subfigure[]{\includegraphics[width=0.99\textwidth]{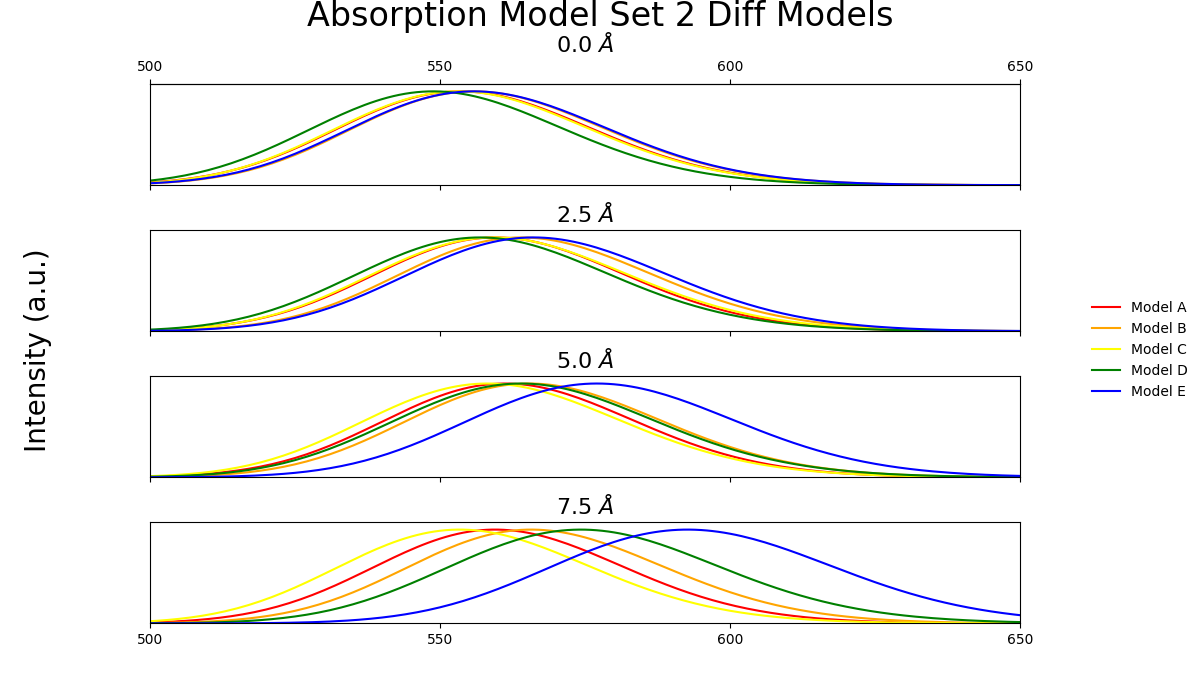}}
    \subfigure[]{\includegraphics[width=0.99\textwidth]{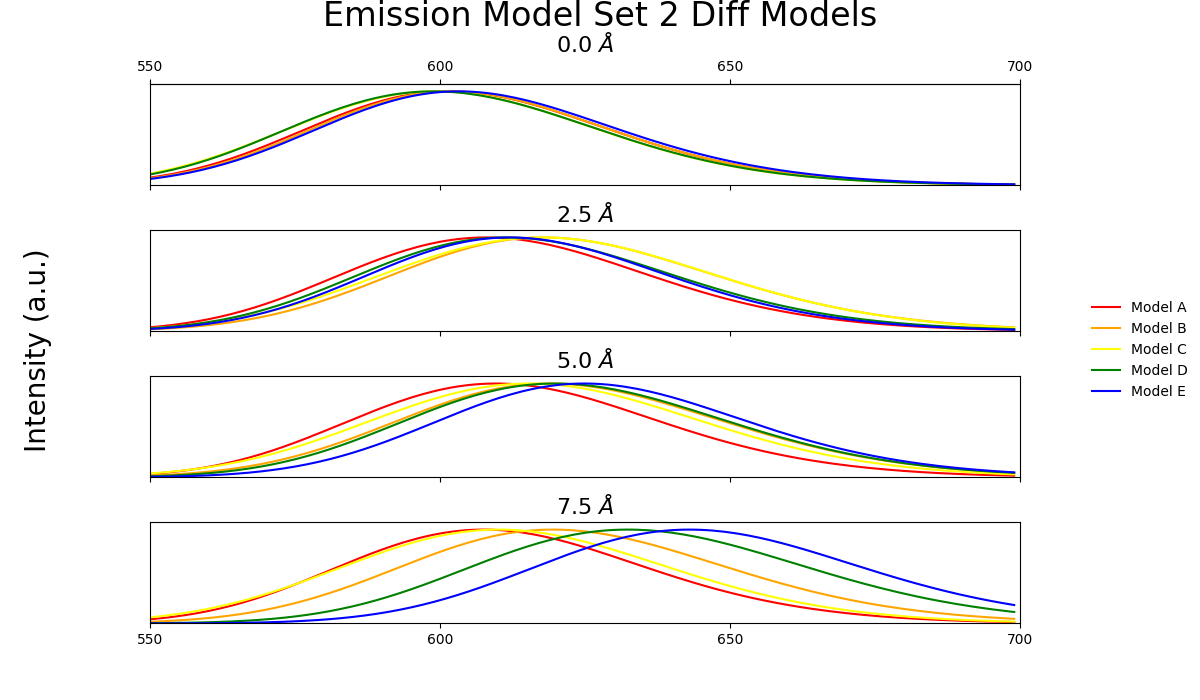}}
    \caption{Vertical excitation spectra predicted using each of the 5 SH-EG models in the committees in model set 2 to directly infer energy gaps for NR in ethanol clusters carved with radii of 0.0, 2.5, 5.0 and 7.5~\AA.}
    \label{fig:committees_over_radii}
\end{figure*}

\begin{figure*}
    \centering
    \subfigure[]{\includegraphics[width=0.49\textwidth]{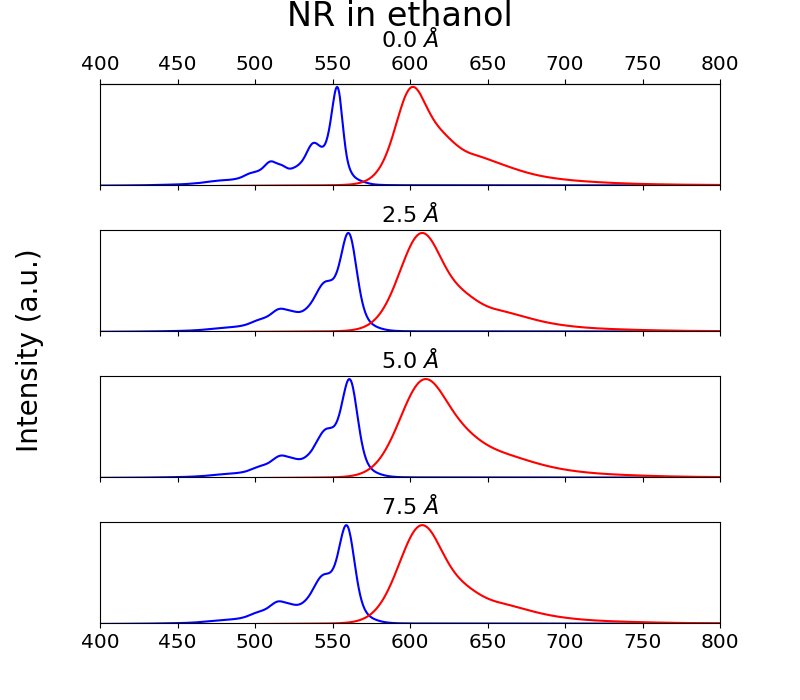}}
    \subfigure[]{\includegraphics[width=0.49\textwidth]{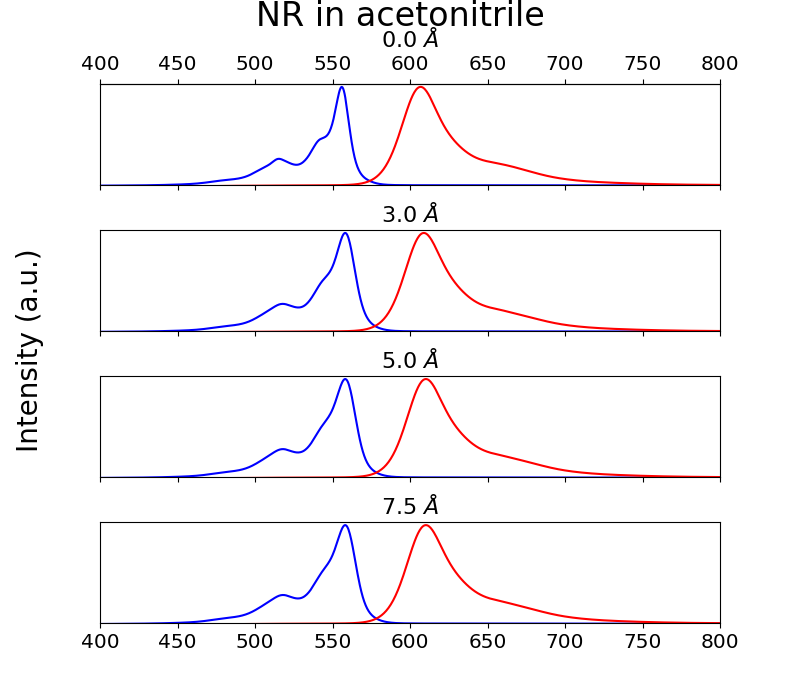}}
    \subfigure[]{\includegraphics[width=0.49\textwidth]{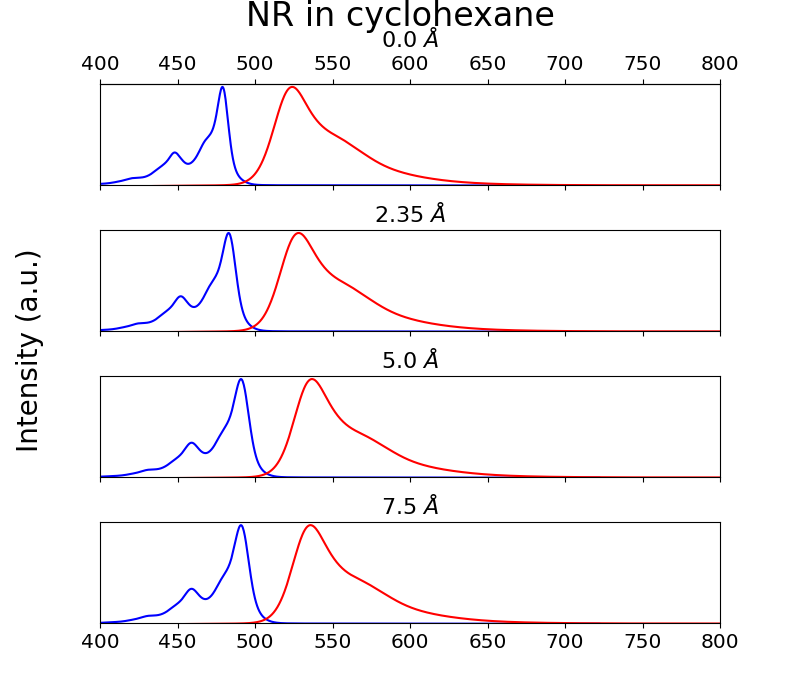}}
    \caption{Absorption and emission spectra generated through the evaluation of clusters of (a) nile red in  ethanol, (b) nile red in acetonitrile and (c) nile red in cyclohexane, with varying radii, using the SH-EG calculators from model set 2.}
    \label{fig:spectra_solvents_radii}
\end{figure*}

\begin{figure*}
    \centering
    \subfigure[]{\includegraphics{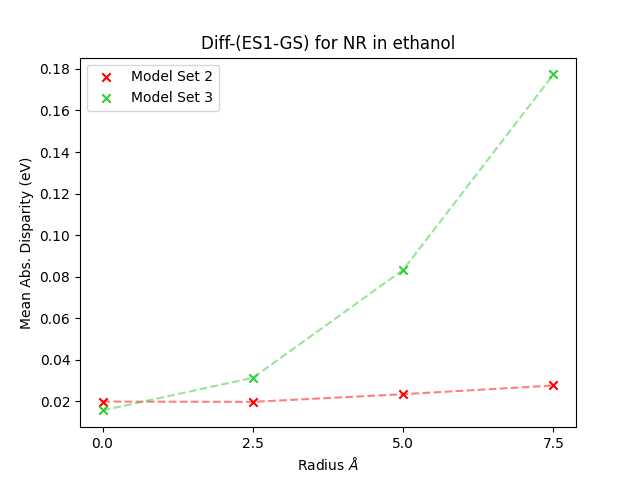}}
    \subfigure[]{\includegraphics[width=0.49\textwidth]{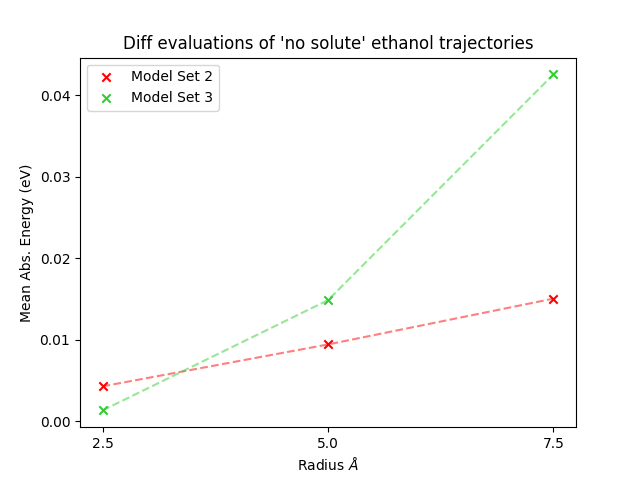}}
    \caption{a.) Mean absolute disparity between energy gaps predicted directly using a EG model/head, and those predicted by subtracting energies predicted by a ground state model/head from an excited state model, for model sets 2 \& 3. The cluster trajectories used for this were the carved production run trajectories used to predict absorption spectra with model set 3. b.) Mean absolute energies predicted by a EG model/head from model sets 2 \& 3 for the carved production run trajectories used to predict absorption spectra with model set 3, augmented by removing NR.}
    \label{fig:sources_of_error_MH}
\end{figure*}

\begin{figure*}
    \centering
    \subfigure[]{\includegraphics[width=0.49\textwidth]{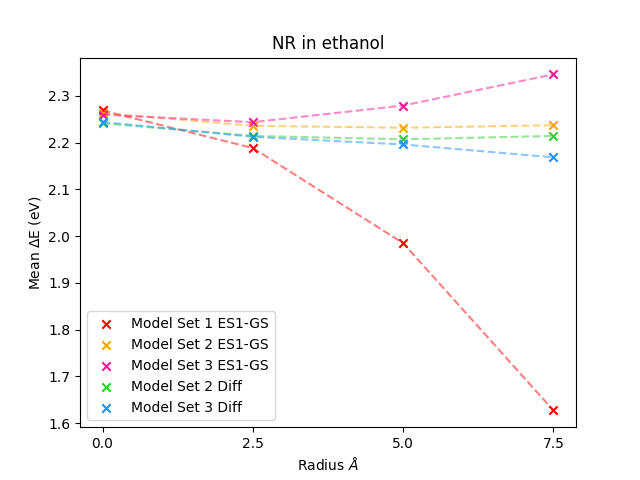}}
    \subfigure[]{\includegraphics[width=0.49\textwidth]{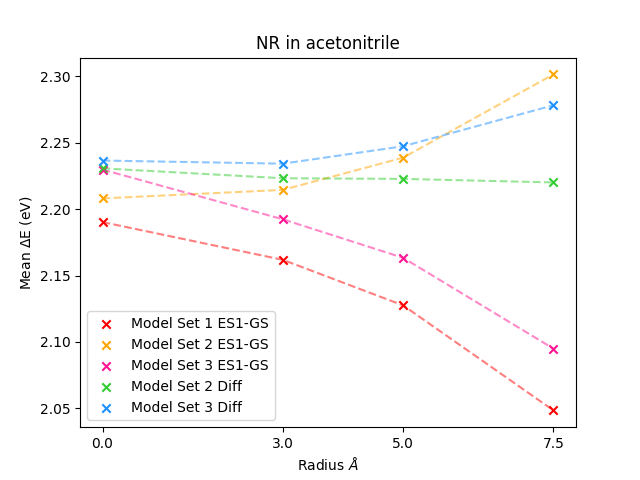}}
    \subfigure[]{\includegraphics[width=0.49\textwidth]{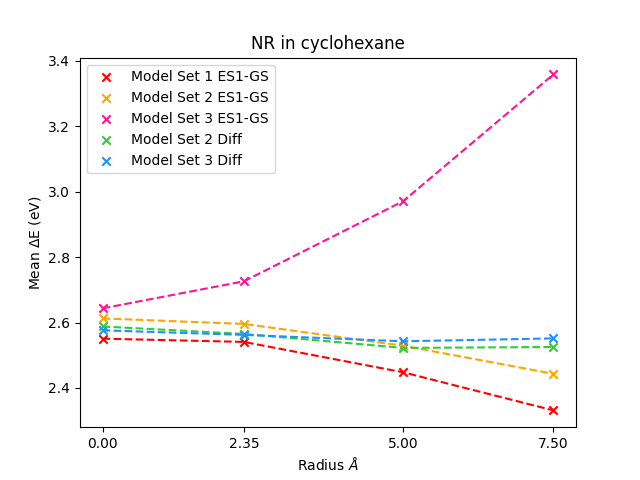}}
    \caption{Mean predicted energy gaps by model sets 1-3 (using one member of each committee), obtained using both methods of energy gap prediction where available, for production run clusters of varying radii.}
    \label{fig:error_models_radius_systems}
\end{figure*}

\begin{figure*}
    \centering
    \includegraphics[width=1\linewidth]{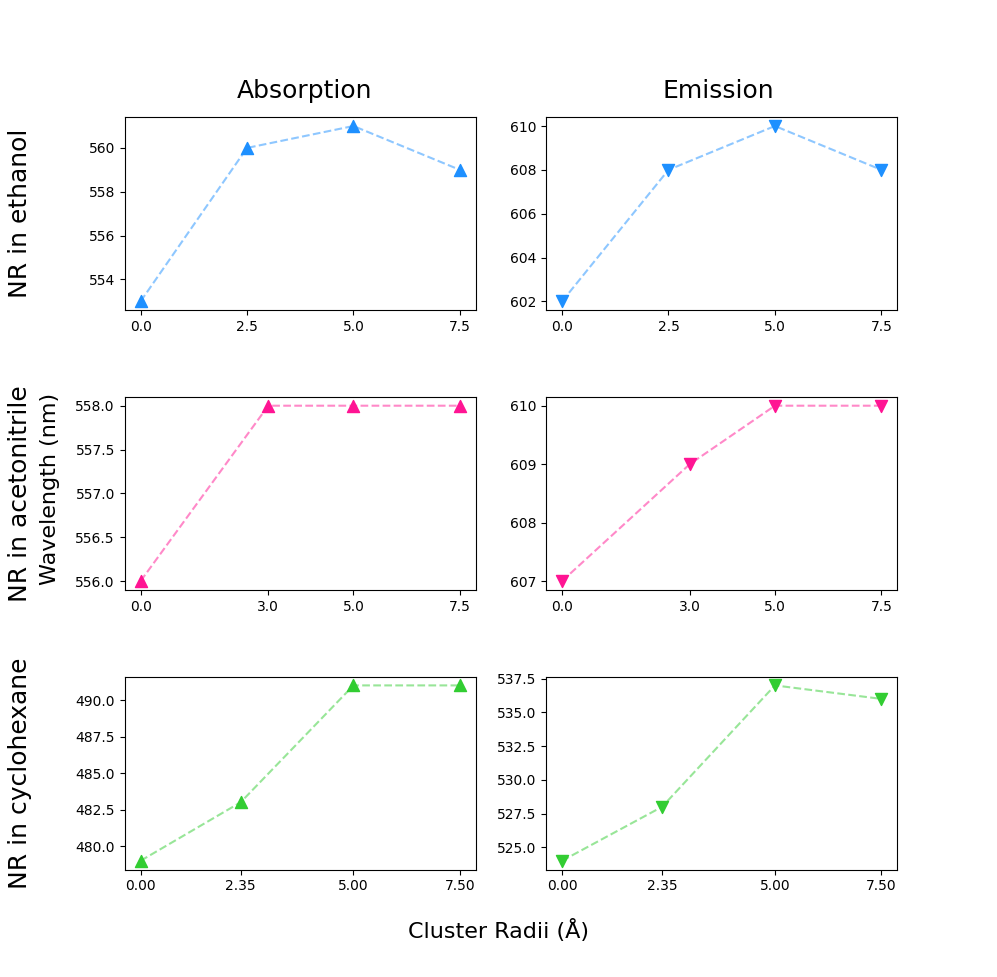}
    \caption{Peak maxima predicted for all three systems using the EG model in model set 2, as a function of production run cluster radii.}
    \label{fig:peakmaxima_radii_systems}
\end{figure*}

\begin{figure*}
    \centering
    \includegraphics[width=1\linewidth]{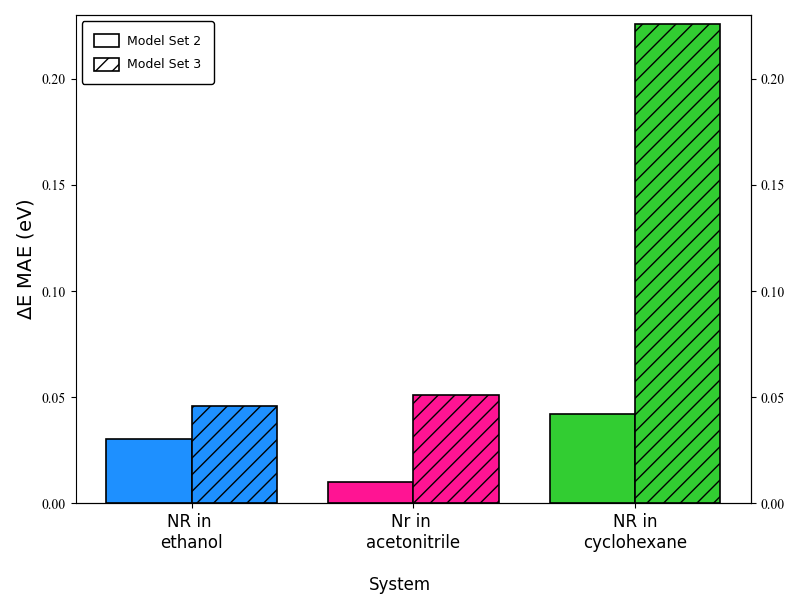}
    \caption{Mean absolute errors in the predicted energy gaps for solute-in-solvent test set 2 obtained by subtracting the ground state model/head energies from excited state model/head energies, for model sets 2 \& 3.}
    \label{fig:errores1gs_sh_mh}
\end{figure*}

\begin{figure*}
    \centering
    \includegraphics[width=1\linewidth]{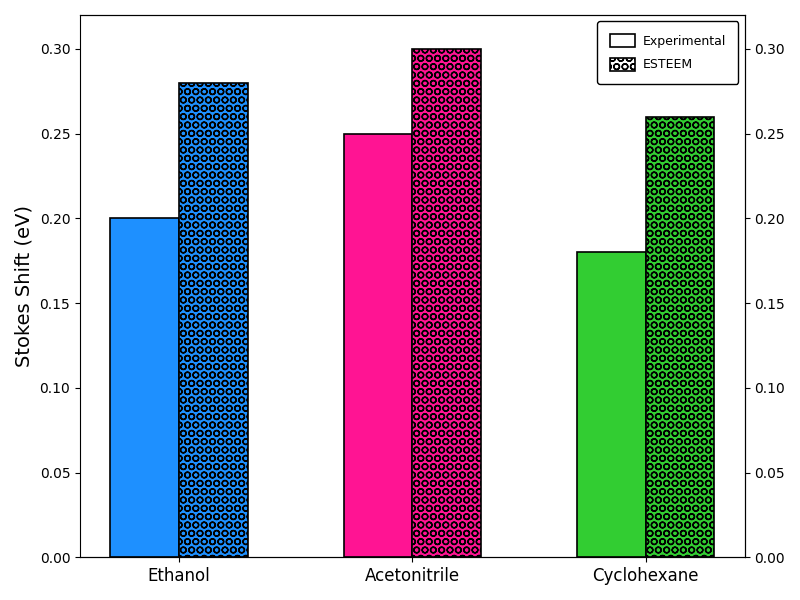}
    \caption{Stokes shifts found in experimental spectra and in spectra generated using model set 2 with a SH-Diff head.}
    \label{fig:stokesshifts_systems}
\end{figure*}

\begin{figure*}
    \centering
    \includegraphics[width=1\linewidth]{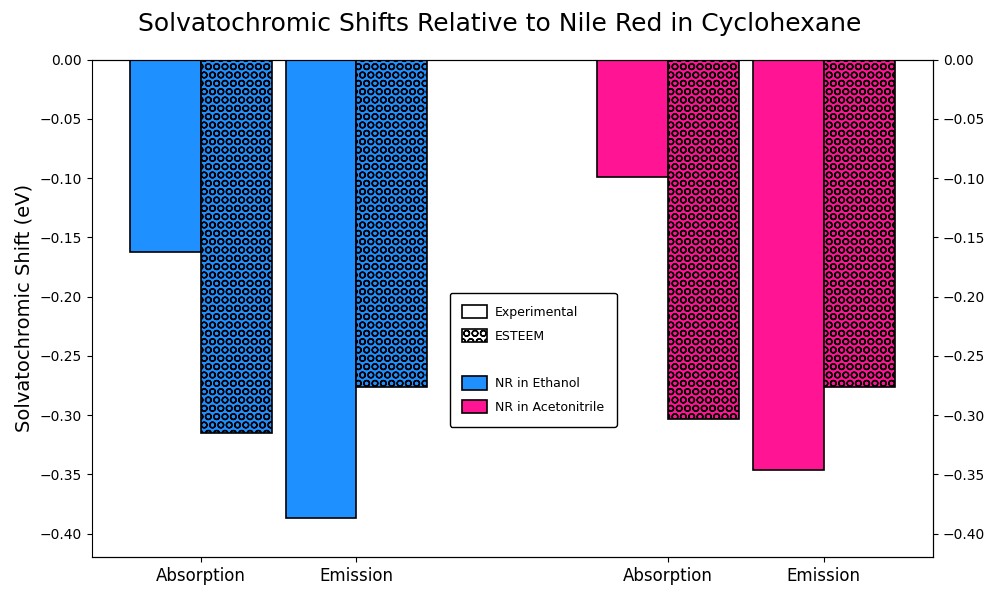}
    \caption{Solvatochromic shifts in peak maxima of NR in ethanol and acetonitrile, relative to the peaks obtained for NR in cyclohexane, found experimentally and using model set 2 with a SH-Diff head.}
    \label{fig:solvatochromicshifts_systems}
\end{figure*}

\FloatBarrier

\subsubsection{SH vs MH model Performance}
To determine which EG models were most effective, we carried out four different tests assessing the performance of both the SH-EG and MH-EG models.

\textbf{No Solute Trajectory Evaluations:} The 2.5~\AA, 5.0~\AA\ and 7.5~\AA\ carved production run trajectories for nile red in ethanol--obtained in section 3.1.3 using the corresponding MH model--were augmented by deleting the nile red molecule from each snapshot. The mean absolute excitation energy associated with the snapshots in each of these augmented trajectories was then predicted using the corresponding SH-EG and MH-EG models. With the chromophore removed from all snapshots, the electronic transition the MLIPs are trained to predict is not possible. Thus, a prediction of 0~eV would indicate that the MLIPs have correctly learned that the presence of the solute is essential.

\textbf{Alignment of EG Calculators with Ground and Excited State Calculators:} The 0.0~\AA, 2.5~\AA, 5.0~\AA\ and 7.5~\AA\ carved trajectories generated in section 3.1.3 for ethanolic nile red were evaluated with a SH-GS, SH-ES1 and SH-EG model (model set 2), as well as all three heads of the corresponding MH from model set 3. Two new sets of trajectories for each carved radius were generated by subtracting the ground state predicted energies from the first excited state energies, for both the SH and MH models. The energies associated with the resulting trajectories were then subtracted from the EG trajectories, to assess their level of agreement in the prediction of excitation energies.

\textbf{Divergence of Excitation Energies with increasing $R_{carve}$:}
The excitation energies associated with each carved production run trajectory generated in section 3.1.3 for all three solutions were predicted in two ways: firstly, by subtracting the ground state energies from the first excited state energies, both inferred by the SH and MH models; and secondly, by evaluating the trajectories with the SH-EG and MH-EG models.

\textbf{Errors in Predicted Excitation Energies for Larger Solution Clusters:}

Carved the final iteration of test clusters to 4.5A, 3.0A and 2.5A, then evaluated with 2nd gen GS and ES + MH ES and GS models. Calculated the average absolute error in energy gap prediction.

$$
\Delta E = E_\mathrm{ES1}-E_\mathrm{GS}
$$

$$
Err_{Solv} = \lvert E^{NS}_\mathrm{ES1} - E^{NS}_\mathrm{GS} \rvert
$$

\end{document}